\documentclass[floatfix, noshowpacs, preprintnumbers, twocolumn, amsmath, amssymb, aps, pra, superscriptaddress, nofootinbib]{revtex4-2}

\usepackage{graphicx, xcolor}
\usepackage{float}
\usepackage{threeparttable}
\usepackage{booktabs}
\usepackage{physics}
\usepackage{makecell}
\usepackage{upquote}
\usepackage{listings}

\definecolor{codegreen}{rgb}{0,0.6,0}
\definecolor{codegray}{rgb}{0.5,0.5,0.5}
\definecolor{codepurple}{rgb}{0.58,0,0.82}
\definecolor{backcolour}{rgb}{0.95,0.95,0.92}

\usepackage{hyperref}

\newcommand{\equalcontrib}{These authors contributed equally to this work.}

\begin{document}
\title{DeepQuantum: A PyTorch-based Software Platform for Quantum Machine Learning and Photonic Quantum Computing}

\author{Jun-Jie He\textsuperscript{$\P$}}
\begingroup
    \renewcommand{\thefootnote}{$\P$}
    \footnotetext{\href{mailto:hejunjie@turingq.com}{hejunjie@turingq.com}}
\endgroup
\thanks{\equalcontrib}
\affiliation{TuringQ Co., Ltd., Shanghai 200240, China}

\author{Ke-Ming Hu}
\thanks{\equalcontrib}
\affiliation{TuringQ Co., Ltd., Shanghai 200240, China}

\author{Yu-Ze Zhu}
\thanks{\equalcontrib}
\affiliation{TuringQ Co., Ltd., Shanghai 200240, China}

\author{Guan-Ju Yan}
\thanks{\equalcontrib}
\affiliation{Center for Integrated Quantum Information Technologies (IQIT), School of Physics and Astronomy and State Key Laboratory of Photonics and Communications, Shanghai Jiao Tong University, Shanghai 200240, China}

\author{Shu-Yi Liang}
\thanks{\equalcontrib}
\affiliation{Center for Integrated Quantum Information Technologies (IQIT), School of Physics and Astronomy and State Key Laboratory of Photonics and Communications, Shanghai Jiao Tong University, Shanghai 200240, China}

\author{Xiang Zhao}
\affiliation{TuringQ Co., Ltd., Shanghai 200240, China}

\author{Ding Wang}
\affiliation{TuringQ Co., Ltd., Shanghai 200240, China}

\author{Fei-Xiang Guo}
\affiliation{Center for Integrated Quantum Information Technologies (IQIT), School of Physics and Astronomy and State Key Laboratory of Photonics and Communications, Shanghai Jiao Tong University, Shanghai 200240, China}
\affiliation{Hefei National Laboratory, Hefei 230088, China}

\author{Ze-Feng Lan}
\affiliation{Center for Integrated Quantum Information Technologies (IQIT), School of Physics and Astronomy and State Key Laboratory of Photonics and Communications, Shanghai Jiao Tong University, Shanghai 200240, China}
\affiliation{Hefei National Laboratory, Hefei 230088, China}

\author{Xiao-Wen Shang}
\affiliation{Center for Integrated Quantum Information Technologies (IQIT), School of Physics and Astronomy and State Key Laboratory of Photonics and Communications, Shanghai Jiao Tong University, Shanghai 200240, China}
\affiliation{Hefei National Laboratory, Hefei 230088, China}

\author{Zi-Ming Yin}
\affiliation{Center for Integrated Quantum Information Technologies (IQIT), School of Physics and Astronomy and State Key Laboratory of Photonics and Communications, Shanghai Jiao Tong University, Shanghai 200240, China}

\author{Xin-Yang Jiang}
\affiliation{Center for Integrated Quantum Information Technologies (IQIT), School of Physics and Astronomy and State Key Laboratory of Photonics and Communications, Shanghai Jiao Tong University, Shanghai 200240, China}

\author{Lin Yang}
\email{yanglin@turingq.com}
\affiliation{TuringQ Co., Ltd., Shanghai 200240, China}

\author{Hao Tang}
\email{htang2015@sjtu.edu.cn}
\affiliation{Center for Integrated Quantum Information Technologies (IQIT), School of Physics and Astronomy and State Key Laboratory of Photonics and Communications, Shanghai Jiao Tong University, Shanghai 200240, China}
\affiliation{Hefei National Laboratory, Hefei 230088, China}

\author{Xian-Min Jin}
\email{xianmin.jin@sjtu.edu.cn}
\affiliation{TuringQ Co., Ltd., Shanghai 200240, China}
\affiliation{Center for Integrated Quantum Information Technologies (IQIT), School of Physics and Astronomy and State Key Laboratory of Photonics and Communications, Shanghai Jiao Tong University, Shanghai 200240, China}
\affiliation{Hefei National Laboratory, Hefei 230088, China}
\affiliation{Chip Hub for Integrated Photonics Xplore (CHIPX), Shanghai Jiao Tong University, Wuxi 214000, China}

\begin{abstract}
We introduce DeepQuantum, an open-source, PyTorch-based software platform for quantum machine learning and photonic quantum computing.
This AI-enhanced framework enables efficient design and execution of hybrid quantum-classical models and variational quantum algorithms on both CPUs and GPUs.
For photonic quantum computing, DeepQuantum implements Fock, Gaussian, and Bosonic backends, catering to different simulation needs.
To our knowledge, it is the first framework to realize closed-loop integration of three paradigms of quantum computing, namely quantum circuits, photonic quantum circuits, and measurement-based quantum computing, thereby enabling robust support for both specialized and universal photonic quantum algorithm design.
Furthermore, DeepQuantum supports large-scale simulations based on tensor network techniques and a distributed parallel computing architecture.
We demonstrate these capabilities through comprehensive benchmarks and illustrative examples.
With its unique features, DeepQuantum is intended to be a powerful platform for both AI for Quantum and Quantum for AI.

\end{abstract}
\maketitle

{
    \hypersetup{hidelinks}
    \tableofcontents
}

\section{Introduction}
Artificial intelligence (AI) and quantum computing (QC), both concepts conceived in the last century, are today at vastly different stages of maturity.
While AI has become a transformative technology driving real-world value across countless domains, QC is in a nascent phase, primarily demonstrating its potential through carefully constructed proofs of quantum advantage.
The pursuit of this advantage, where quantum devices perform tasks intractable for the most powerful classical supercomputers, has seen remarkable progress in recent years.
Landmark experiments, such as quantum random sampling on superconducting and photonic platforms~\cite{arute2019quantum, wu2021strong, zhong2020quantum}, have convincingly demonstrated this capability for specific, contrived problems.
However, a critical and pressing challenge for the field is to translate these theoretical demonstrations into practical, real-world applications with tangible value.
The transition from demonstrating quantum advantage to harnessing quantum utility is of paramount importance.

In the current noisy intermediate-scale quantum (NISQ) era~\cite{preskill2018quantum}, this translation effort relies heavily on the synergy between quantum processors and classical computers.
Hybrid quantum-classical algorithms have emerged as the most promising paradigm~\cite{callison2022hybrid}, where a quantum device acts as a specialized co-processor whose parameters are optimized within a classical post-processing and feed-forward loop.
Prominent examples that exemplify this approach include the variational quantum eigensolver (VQE) for quantum chemistry~\cite{peruzzo2014variational}, the quantum approximate optimization algorithm (QAOA) for combinatorial optimization~\cite{farhi2014quantum}, and a variety of methods in quantum machine learning (QML), such as quantum reservoir computing and quantum extreme learning machines~\cite{ghosh2019quantum, mujal2021opportunities}.
These works demonstrate how quantum circuits can be utilized as powerful models within classical AI frameworks to achieve a ``quantum-enhanced'' effect~\cite{dunjko2016quantum, havlivcek2019supervised}.

As the community looks beyond the NISQ era, the ultimate goal is the realization of fault-tolerant quantum computing (FTQC), which promises to unlock the full potential of seminal algorithms like Shor's~\cite{shor1999polynomial} and Grover's~\cite{grover1996fast}.
The journey from noisy, intermediate-scale devices to robust, large-scale fault-tolerant machines is fraught with immense challenges, particularly in areas including quantum error correction (QEC) and circuit compilation.
Increasingly, AI is being recognized as an indispensable tool to overcome these hurdles.
For instance, machine learning, especially deep reinforcement learning, has shown significant promise in designing more efficient QEC decoders~\cite{andreasson2019quantum, nautrup2019optimizing} and optimizing the complex process of compiling high-level quantum algorithms into low-level hardware instructions~\cite{fosel2021quantum, pozzi2022using}.

As established, AI is crucial for advancing QC, from optimizing NISQ-era utilities to paving the way for FTQC.
Conversely, the nascent field of QML provides compelling evidence for the reverse.
Early studies suggest that quantum algorithms could enhance AI by offering faster training, improved model performance, or reduced data requirements for specific tasks~\cite{acampora2025quantum}.
This creates a powerful virtuous cycle: AI accelerates the development of quantum hardware, which in turn promises to revolutionize AI itself.
This reciprocal relationship underscores a future where the two fields are deeply intertwined.

Given this profound and growing interdependency, the need for a programming framework that deeply and seamlessly integrates AI and QC is paramount.
Several pioneering frameworks have emerged to address this, such as PennyLane~\cite{bergholm2018pennylane} and TensorFlow Quantum~\cite{broughton2020tensorflow}, which successfully treat quantum circuits as differentiable components within classical machine learning pipelines.
While these tools have significantly advanced the field of QML, researchers may face practical challenges when attempting to couple quantum circuits with the ever-evolving landscape of state-of-the-art AI models.
Furthermore, they predominantly focus on the standard qubit-based circuit model.
However, to pave a viable path toward FTQC, it is crucial to consider alternative computational models and hardware platforms.
Photonic quantum computing, in particular, stands out as a leading candidate due to its inherent advantages, such as low noise profiles, operational simplicity, and its role as a natural ``flying qubit'' for networking.
This versatility has spurred recent breakthroughs, including the generation of integrated photonic Gottesman-Kitaev-Preskill (GKP) qubits~\cite{larsen2025integrated} and demonstrations of modular, manufacturable computing platforms~\cite{aghaee2025scaling, psiquantum2025manufacturable}.
Many of these scalable architectures are designed around the measurement-based quantum computing (MBQC) paradigm~\cite{raussendorf2001one}, which is naturally suited for fault-tolerant computation.
Existing specialized frameworks, such as Strawberry Fields~\cite{killoran2019strawberry} and Piquasso~\cite{kolarovszki2025piquasso} for continuous-variable (CV) quantum computation, Perceval~\cite{heurtel2023perceval} for discrete-variable (DV) photonic systems, and Graphix~\cite{sunami2022graphix, uldemolins2026graphix} for MBQC simulations, have provided invaluable tools for their respective domains, but a unified platform that bridges these models while maintaining a native AI integration is still lacking.

To address these challenges and bridge the existing gaps, we introduce DeepQuantum, a novel programming framework engineered to foster the deep synergy between AI and QC for the NISQ era and beyond.
This paper is organized as follows: Section~\ref{sec:overview} introduces the DeepQuantum framework, highlighting its global architecture and key features.
Section~\ref{sec:benchmark} validates such advantages through benchmark results.
Sections~\ref{sec:qubit},~\ref{sec:qumode},~and~\ref{sec:mbqc} respectively focus on quantum computation with qubits, photonic qumodes, and MBQC patterns, covering the background, API overview, and application examples.
Section~\ref{sec:large-scale} demonstrates large-scale quantum simulation through tensor network techniques and distributed methods.
In Section~\ref{sec:conclusion}, we summarize the whole paper and outline the future outlook.

\section{DeepQuantum Overview}
\label{sec:overview}
The DeepQuantum platform is designed to bridge AI and QC, providing an integrated environment that supports both AI-enhanced quantum algorithms and quantum-enhanced machine learning.
Built with an emphasis on efficiency and usability, it offers a high-performance programming framework and an intuitive interface for constructing quantum circuits, allowing developers to prototype and explore advanced quantum applications rapidly.

Implemented in Python and leveraging the widely adopted deep learning framework PyTorch~\cite{paszke2019pytorch}, DeepQuantum is fully open-source\footnote[1]{\url{https://github.com/TuringQ/deepquantum}} and can be easily installed via pip:

\begin{lstlisting}[
    language=bash,
    numbers=none,
    backgroundcolor={\color[gray]{0.95}},
    basicstyle=\ttfamily\small
]
$ pip install deepquantum
\end{lstlisting}

\begin{figure*}[!htbp]
    \centering
    \includegraphics[width=\linewidth]{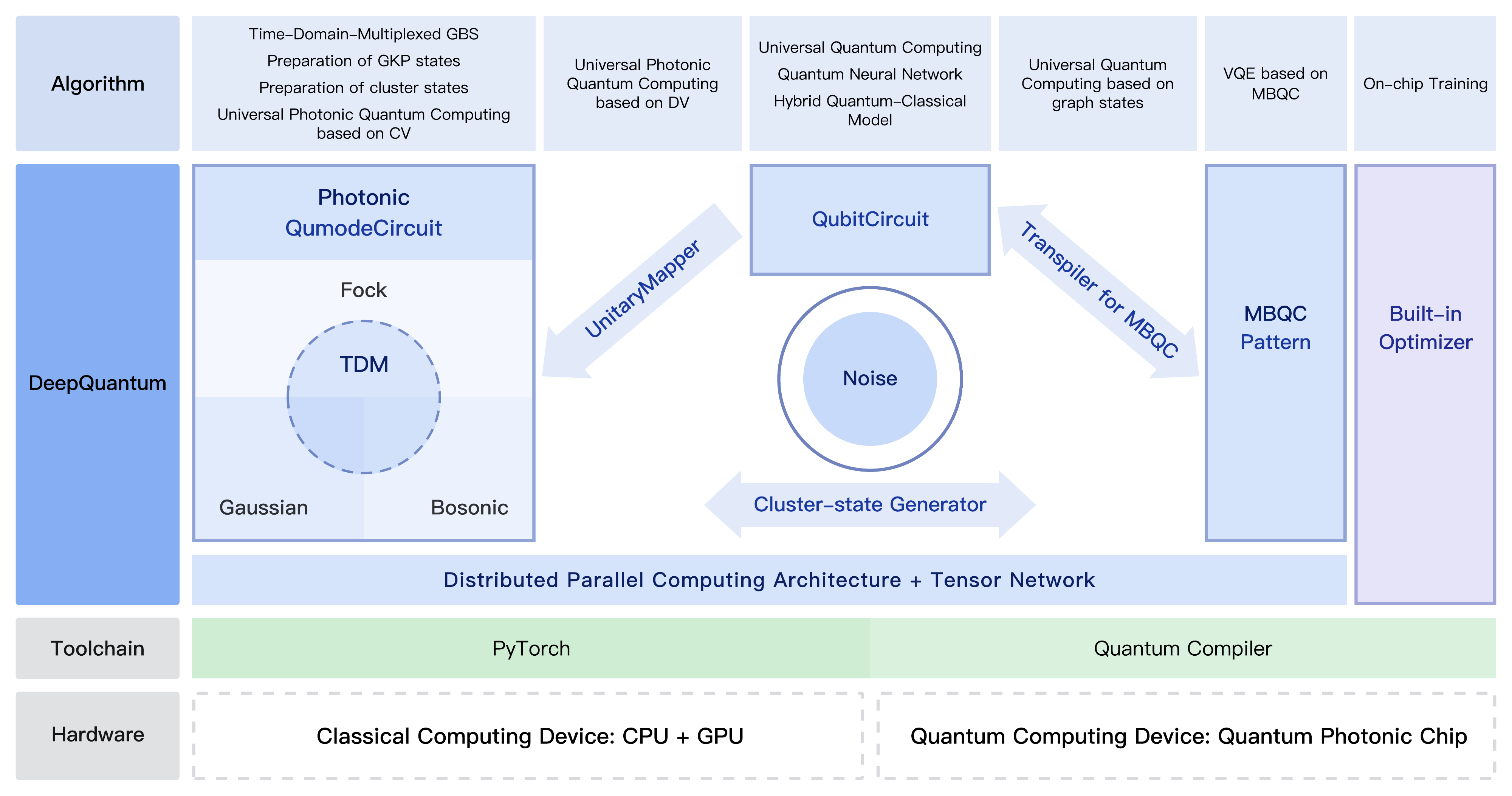}
    \caption{The architecture of DeepQuantum and its ecosystem, including high-level quantum algorithms, toolchain, and low-level hardware.}
    \label{fig:architecture}
\end{figure*}

The overall architecture of DeepQuantum is illustrated in Fig.~\ref{fig:architecture}.
As an efficient programming framework tailored for quantum machine learning and photonic quantum computing, it achieves a closed-loop integration of three major paradigms: quantum circuits, photonic quantum circuits, and MBQC.
At its core, the framework provides three principal classes for quantum computation: \texttt{QubitCircuit}, \texttt{QumodeCircuit}, and \texttt{Pattern}.
The \texttt{QubitCircuit} class enables users to easily construct and simulate quantum circuits, facilitating the rapid design and optimization of quantum neural networks.
The \texttt{QumodeCircuit} class includes Fock, Gaussian, and Bosonic backends, allowing in-depth exploration of photonic quantum circuits.
Furthermore, \texttt{QumodeCircuit} supports the development of practical applications based on time-domain multiplexing (TDM) and Gaussian Boson sampling (GBS) algorithms~\cite{hamilton2017gaussian}.
The \texttt{Pattern} class, along with the built-in transpiler from \texttt{QubitCircuit} to \texttt{Pattern}, empowers users to extensively investigate the MBQC paradigm.
Theoretically, these three paradigms are closely linked and interconvertible.
MBQC has been proven to be equivalent to the quantum circuit model~\cite{raussendorf2001one}.
Photonic quantum circuits naturally support the generation of cluster states through appropriate algorithms.
Specifically, DeepQuantum provides \texttt{UnitaryMapper} to help transfer \texttt{QubitCircuit} to \texttt{QumodeCircuit} based on dual-rail encoding.
DeepQuantum also supports noisy simulations with built-in noise channels.
To support large-scale simulations, DeepQuantum incorporates a distributed parallel computing architecture and leverages tensor network techniques.

Fig.~\ref{fig:architecture} also showcases the central role of DeepQuantum in the ecosystem and as a bridge between high-level quantum algorithms and low-level hardware execution.
Through its built-in modules, DeepQuantum can handle universal quantum computing based on qubits, DV, CV, and graph states (see more details in later sections).
On classical hardware (CPUs and GPUs), it utilizes PyTorch's intuitive API and extensive machine learning libraries to facilitate efficient development and experimentation with hybrid quantum-classical algorithms such as QML, QAOA, and VQE.
For execution on photonic quantum processors, DeepQuantum provides built-in gradient-free optimizers to enable on-chip training of QML models.
In brief, DeepQuantum connects user-facing algorithm design to chip-level physical mapping.

DeepQuantum is built upon the principles of conciseness, efficiency, flexibility, and versatility.
Its key features include:

\begin{itemize}
    \item \textbf{AI-Enhanced Quantum Computing Framework}:
    Seamlessly integrated with PyTorch, it utilizes technologies such as automatic differentiation, vectorized parallelism, and GPU acceleration for efficiency.
    It facilitates the easy construction of hybrid quantum-classical models, enabling end-to-end training and inference.
    \item \textbf{User-Friendly API Design}:
    The API is designed to be simple and intuitive, making it easy to initialize quantum neural networks and providing flexibility in data encoding.
    \item \textbf{Photonic Quantum Computing Simulation}:
    The Photonic module includes Fock, Gaussian, and Bosonic backends, catering to different simulation needs in photonic quantum computing.
    It comes with built-in optimizers to support on-chip training of photonic quantum circuits.
    \item \textbf{Large-Scale Quantum Circuit Simulation}:
    Leveraging tensor network techniques, DeepQuantum enables approximate simulation of circuits with over 100 qubits on a single laptop under favorable entanglement and bond-dimension conditions.
    Through a distributed parallel architecture and PyTorch's native communication protocol, it efficiently utilizes multi-node, multi-GPU computational power to boost large-scale quantum simulations.
    These capabilities allow DeepQuantum to deliver industry-leading performance in both simulation scale and efficiency.
    \item \textbf{Advanced Architecture for Cutting-Edge Algorithm Exploration}:
    It is the first framework to support algorithm design and mapping for TDM photonic quantum circuits, and the first to realize closed-loop integration of quantum circuits, photonic quantum circuits, and MBQC, enabling robust support for both specialized and universal photonic quantum algorithm design.
\end{itemize}

\section{Benchmarks}
\label{sec:benchmark}
DeepQuantum is engineered for high-performance quantum simulation, utilizing optimized data structures to enhance runtime efficiency.
In this section, we present comprehensive benchmarks comparing DeepQuantum against several established quantum computing programming frameworks.
The evaluation covers core functionalities, including gradient and Hessian calculations for \texttt{QubitCircuit}, the computation of Hafnians, Torontonians, and permanents for \texttt{QumodeCircuit}, as well as transpilation and forward simulation for \texttt{Pattern}.
Each benchmark is executed ten times following an initial warm-up iteration.
Note that for computationally demanding configurations, the number of repetitions is reduced to ensure a reasonable overall runtime.

\begin{table}[!htbp]
    \caption{Software versions of the benchmarking frameworks.}
    \label{table:framework_version}
    \centering
    \tabcolsep=0.4cm
    \begin{tabular}{cc}
        \toprule[1.5pt]
        \textbf{Framework}                              & \textbf{Version} \\
        \midrule[1pt]
        DeepQuantum                                     & 4.5.0            \\
        PennyLane~\cite{bergholm2018pennylane}          & 0.41.1           \\
        MindQuantum~\cite{xu2024mindspore}              & 0.10.0           \\
        Strawberry Fields~\cite{killoran2019strawberry} & 0.23.0           \\
        The Walrus~\cite{gupt2019walrus}                & 0.22.0           \\
        Piquasso~\cite{kolarovszki2025piquasso}         & 6.0.1            \\
        Graphix~\cite{sunami2022graphix}                & 0.3.1            \\
        PyQPanda3~\cite{zou2025qpanda3}                 & 0.3.2            \\
        \bottomrule[1.5pt]
    \end{tabular}
\end{table}

\begin{table}[!htbp]
    \caption{Software and hardware environment specifications.}
    \label{table:environment}
    \centering
    \tabcolsep=0.4cm
    \begin{tabular}{cc}
        \toprule[1.5pt]
        \textbf{Software/Hardware} & \textbf{Version/Model} \\
        \midrule[1pt]
        PyTorch                    & 2.3.0 + CUDA 12.1      \\
        NumPy                      & 1.26.4                 \\
        CPU                        & Intel Xeon Gold 6326   \\
        GPU                        & NVIDIA A100            \\
        \bottomrule[1.5pt]
    \end{tabular}
\end{table}

Tables~\ref{table:framework_version} and~\ref{table:environment} summarize the framework versions and the hardware/software specifications utilized in these experiments, respectively.

\subsection{Gradient and Hessian}
\begin{figure}[!htbp]
    \centering
    \includegraphics[width=\linewidth]{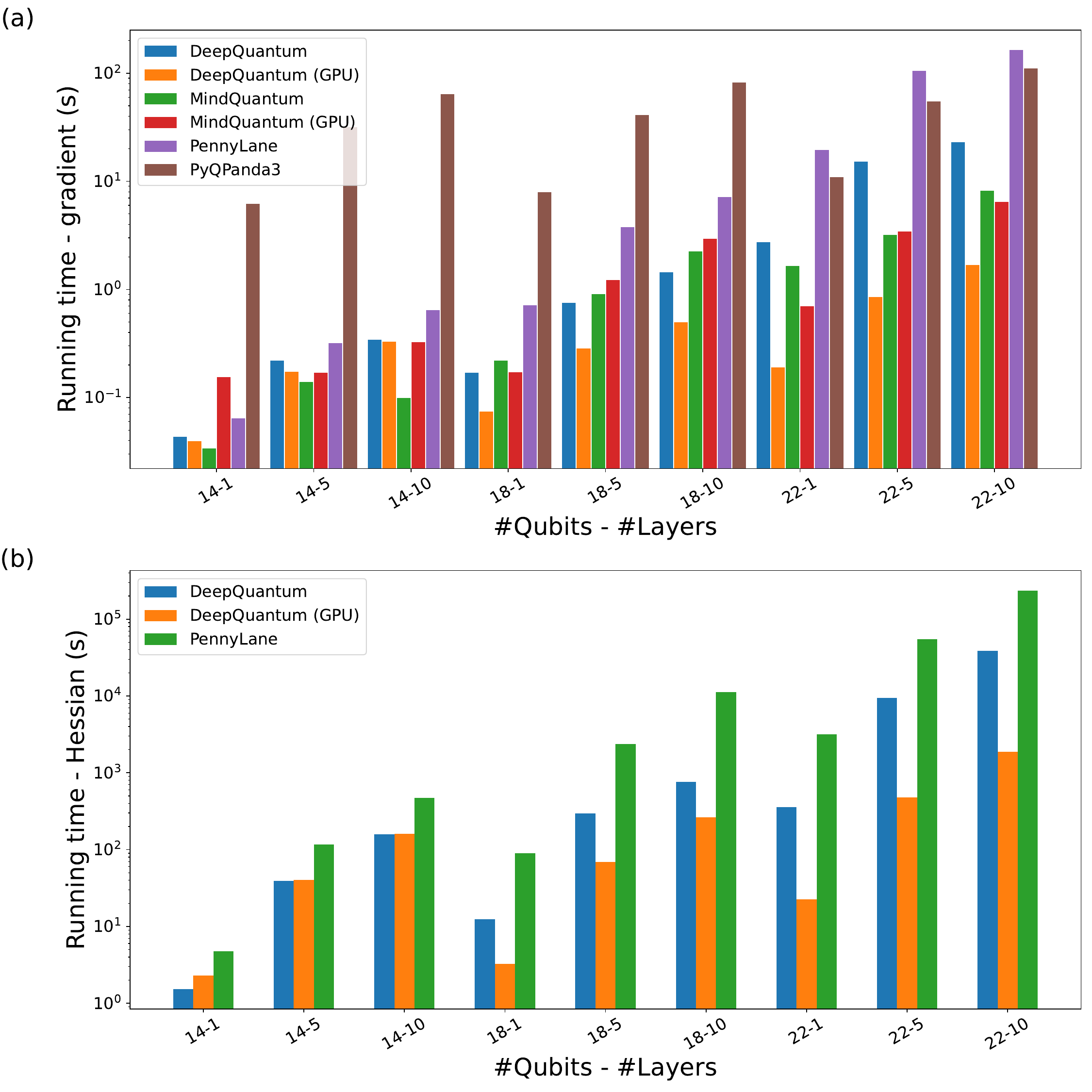}
    \caption{Benchmark results of gradient (a) and Hessian (b) computations.
    Comparison between DeepQuantum, MindQuantum, PennyLane, and PyQPanda3 across varying qubit counts and circuit depths (\#Layers).}
    \label{fig:benchmark_grad_hess}
\end{figure}

Efficient gradient and Hessian computations are pivotal for advancing variational quantum algorithms (VQAs), such as VQE and QAOA~\cite{cerezo2021variational}.
Fig.~\ref{fig:benchmark_grad_hess}(a) illustrates the benchmark results for gradient computation across DeepQuantum, MindQuantum, PennyLane, and PyQPanda3, with varying qubit counts and numbers of layers.

At larger scales, DeepQuantum demonstrates an order-of-magnitude improvement in efficiency over PennyLane and PyQPanda3.
DeepQuantum also outperforms MindQuantum when scaling to 18 qubits.
Regarding GPU acceleration, DeepQuantum (GPU) consistently surpasses CPU-based implementations starting from 14 qubits, ultimately achieving a speedup exceeding one order of magnitude.
Beyond 18 qubits, DeepQuantum (GPU) emerges as the fastest among all tested frameworks.

Fig.~\ref{fig:benchmark_grad_hess}(b) compares the Hessian computation performance between DeepQuantum and PennyLane under identical circuit configurations.
The results indicate that DeepQuantum achieves significantly higher computational speeds.
This performance gap widens at larger scales, with the GPU-accelerated version yielding an approximately 100-fold speedup compared to PennyLane for the 22-qubit configurations.

\subsection{Hafnian, Torontonian and Permanent}
The evaluation of matrix functions, such as the Hafnian~\cite{barvinok2016combinatorics}, Torontonian~\cite{quesada2018gaussian}, and permanent~\cite{valiant1979complexity}, is fundamental to photonic quantum simulation, as these functions determine output probability distributions in GBS (Gaussian backend) or DV linear-optical (Fock backend) models.
While these computations are generally \#P-hard and classically intractable for large-scale problems, efficient evaluation remains essential for verification and small-to-medium-scale simulations in hybrid quantum-classical workflows.

DeepQuantum implements a batch-processing scheme to compute these functions.
This approach enables the simultaneous evaluation of multiple circuit instances, significantly improving throughput and resource utilization on GPUs and multi-core systems.
This is particularly advantageous for AI-driven tasks, such as training variational photonic circuits or running large-scale parameter scans, where parallel processing effectively amortizes the individual computational overhead.

It is worth noting that our current implementation does not incorporate low-level, architecture-specific optimizations for these matrix functions.
Consequently, in single-instance computations, DeepQuantum may trail behind specialized libraries such as Strawberry Fields (via The Walrus) or Piquasso, which rely on highly optimized numerical backends.
However, DeepQuantum's true strength lies in scenarios prioritizing high throughput.

\begin{figure}[!htbp]
    \centering
    \includegraphics[width=\linewidth]{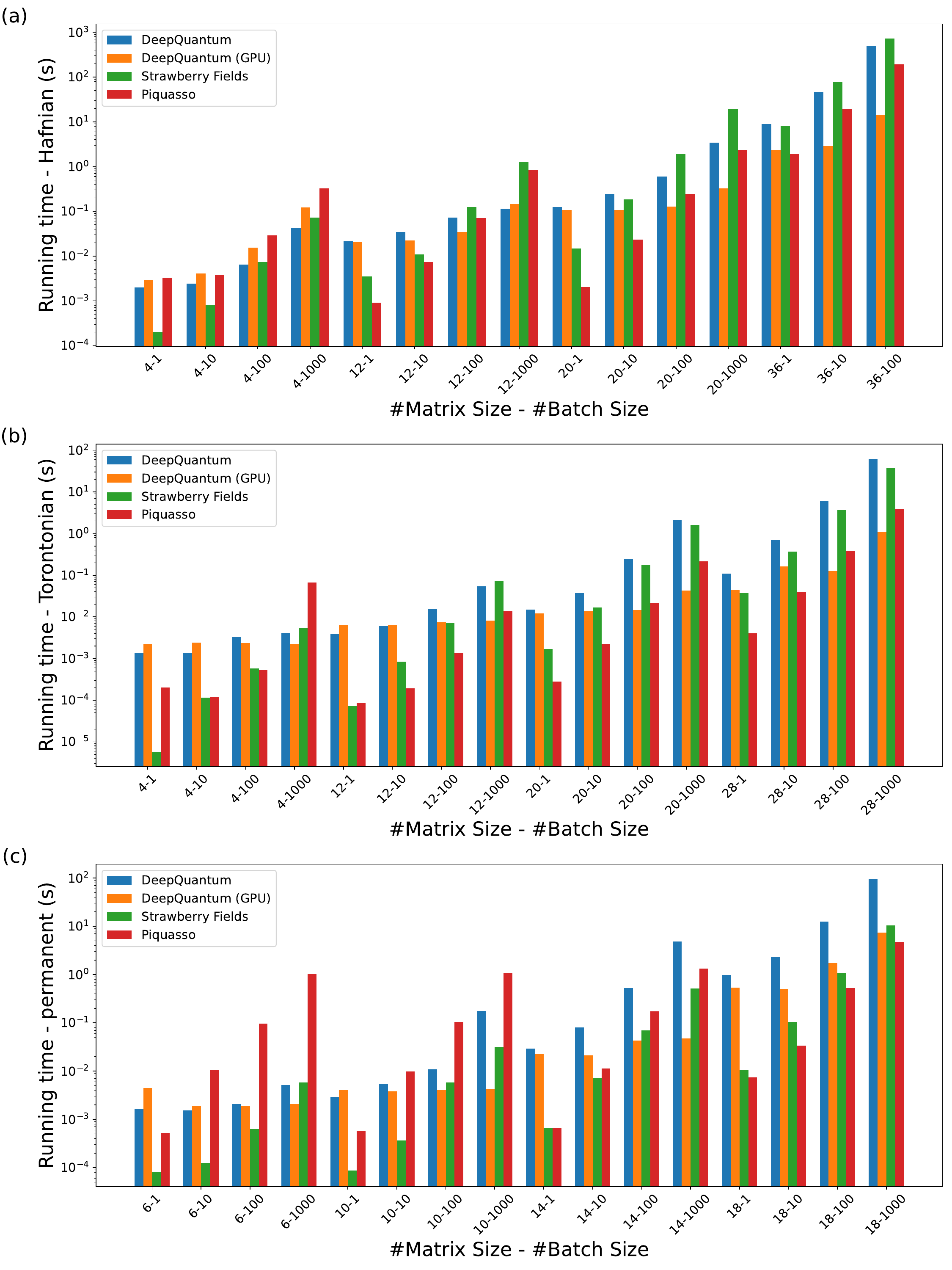}
    \caption{Benchmark results of Hafnian (a), Torontonian (b), and permanent (c) computations.
    Comparison among DeepQuantum, Strawberry Fields, and Piquasso using various matrix and batch sizes.}
    \label{fig:benchmark_haf_tor_per}
\end{figure}

Fig.~\ref{fig:benchmark_haf_tor_per} presents the benchmarking outcomes across different matrix and batch sizes.
As shown in panels (a) and (b), DeepQuantum demonstrates highly competitive performance for Hafnian and Torontonian computations, particularly when handling large batch sizes where other frameworks lack native batch-processing support.
Notably, DeepQuantum (GPU) achieves a significant performance boost, outperforming Strawberry Fields and Piquasso as both batch size and matrix size increase.

Fig.~\ref{fig:benchmark_haf_tor_per}(c) illustrates the performance for permanent computation.
For small-scale systems ($\le 14$ modes), Strawberry Fields achieves the highest speed, whereas Piquasso takes the lead for larger systems ($\ge 18$ modes).
Nevertheless, DeepQuantum (GPU) exhibits minimal sensitivity to batch size scaling, providing enhanced stability and performance in medium-scale applications with large batch sizes (e.g., 10-1000 and 14-1000). This underscores DeepQuantum's proficiency in managing large-scale batch processing.

\subsection{Transpilation and Forward Computation}
We further benchmarked the performance of DeepQuantum against Graphix, an open-source MBQC software package, focusing on two representative tasks: pattern transpilation and forward computation~\cite{sunami2022graphix}.

\begin{figure}[!htbp]
    \centering
    \includegraphics[width=\linewidth]{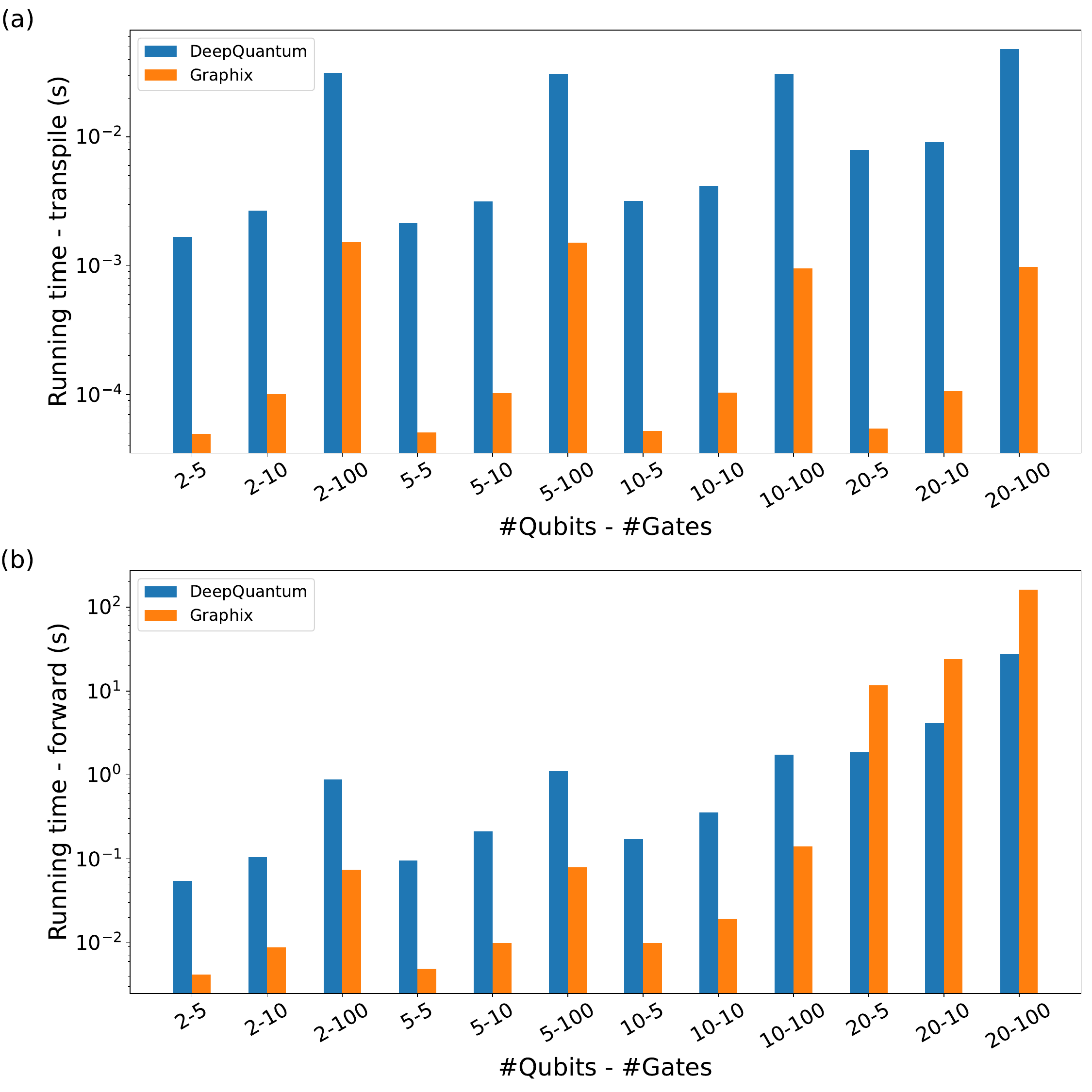}
    \caption{Benchmark results of MBQC pattern transpilation (a) and forward simulation (b).
    Comparison between DeepQuantum and Graphix across different qubit counts and gate numbers.}
    \label{fig:benchmark_mbqc}
\end{figure}

Fig.~\ref{fig:benchmark_mbqc}(a) displays the running time for pattern transpilation under different numbers of qubits and gates.
While Graphix performs this specific step faster than DeepQuantum, the absolute processing time for both frameworks remains extremely short (under $10^{-1}$ seconds).
Therefore, this minor computational overhead can be regarded as negligible within the end-to-end workflow.

In contrast, Fig.~\ref{fig:benchmark_mbqc}(b) compares the much more resource-intensive forward computation of state vectors under identical configurations.
While Graphix exhibits a slight edge for shallow and small-scale patterns, DeepQuantum demonstrates superior scaling efficiency.
For circuits exceeding 20 qubits, DeepQuantum establishes a substantial performance advantage, underscoring its effectiveness in simulating complex MBQC patterns.

\section{Quantum Computing with Qubits}
\label{sec:qubit}
\subsection{Background}
Quantum computing leverages the fundamental principles of quantum mechanics, such as superposition and entanglement, offering the potential to solve specific computational tasks that are intractable for classical computers~\cite{nielsen2010quantum, preskill2018quantum}.
At the core of this paradigm is the quantum bit, or qubit, which serves as the basic unit of quantum information.
An arbitrary pure state of a single qubit can be visualized as a point on the Bloch sphere, parameterized by the spherical coordinates $\theta$ and $\phi$:
\begin{equation}
    \ket{\psi(\theta,\phi)} = \cos\frac{\theta}{2}\ket{0} + e^{i\phi}\sin\frac{\theta}{2}\ket{1},
\end{equation}
where $\ket{0} \equiv \begin{bmatrix} 1 \\ 0 \end{bmatrix}$ and $\ket{1} \equiv \begin{bmatrix} 0 \\ 1 \end{bmatrix}$ denote the computational basis states, corresponding to the eigenstates of the Pauli-$Z$ operator.

To accommodate mixed states and statistical ensembles, quantum systems are more comprehensively described using density matrices.
In a given orthonormal basis $\{\ket{i}\}$, a density matrix $\rho$ is characterized by its matrix elements:
\begin{equation}
    \rho_{ij} \equiv \mel{i}{\rho}{j}.
\end{equation}

Transformations on qubits are performed via quantum gates, which are mathematically represented by unitary operators $\hat{U}$.
A quantum computation typically proceeds by applying a sequence of such gates to an initial state.
For a pure state vector, the discrete time evolution is given by
\begin{equation}
    \ket{\psi_n} = \hat{U}_n \cdots \hat{U}_1 \ket{\psi_0}.
\end{equation}
Equivalently, in the density matrix formalism, this evolution is expressed as
\begin{equation}
    \rho_n = \hat{U}_n \cdots \hat{U}_1 \rho_0 \hat{U}_1^\dagger \cdots \hat{U}_n^\dagger,
\end{equation}
where $\rho_0 = \ketbra{\psi_0}{\psi_0}$ is the initial density matrix, and $\hat{U}_i$ denotes the $i$-th applied unitary gate.

\subsection{API Overview}
In DeepQuantum, \texttt{QubitCircuit} serves as the core API for simulating quantum circuits.
The entire workflow encompasses the initialization, construction, execution, and measurement of quantum circuits, involving the following four components:

\begin{itemize}
    \item \textbf{Quantum state.}
    Quantum states are represented by the \texttt{QubitState} class, which supports initialization from explicit state vectors or density matrices.
    Internally, states are stored as \texttt{torch.Tensor} objects, enabling native GPU acceleration and seamless integration with PyTorch-based workflows.

    \item \textbf{Quantum operation.}
    DeepQuantum provides a comprehensive library of quantum operations, including all common fixed and parameterized quantum gates, as well as single-qubit noise channels.
    Users can construct a quantum circuit by sequentially adding these operations to a \texttt{QubitCircuit} instance, specifying the target qubits and corresponding parameters.

    \item \textbf{Quantum circuit.}
    The \texttt{QubitCircuit} class acts as the central abstraction for simulation.
    A circuit is initialized with a defined number of qubits and, optionally, a specific initial quantum state.
    Quantum operations can then be appended sequentially to define the computational process.
    Forward execution (invoked via \texttt{QubitCircuit.forward()}) propagates the state through the defined sequence of operations to yield the final evolved state.
    Additionally, DeepQuantum offers built-in functions for circuit visualization, compilation, and advanced representations such as matrix product states (MPS)~\cite{orus2014practical}.

    \item \textbf{Measurement.}
    To extract information from the evolved states, DeepQuantum supports sampling-based measurement via the \texttt{QubitCircuit.measure()} method, and expectation value evaluation via the \texttt{QubitCircuit.expectation()} method.
    Observables must be predefined and appended to the circuit prior to execution using the \texttt{QubitCircuit.observable()} method.
\end{itemize}

Collectively, these components form a unified and end-to-end differentiable framework that allows users to construct quantum algorithms flexibly and efficiently.
Since variational parameters can be natively encoded into a \texttt{QubitCircuit}, this architecture is exceptionally well-suited for VQAs and other hybrid quantum-classical workflows.

\begin{lstlisting}[caption={An illustrative example of \texttt{QubitCircuit}.}, label={code:qubitcircuit}]
import deepquantum as dq
import torch

# Prepare the encoded data
data = torch.tensor([1., 2., 3.])
# Initialize a 3-qubit circuit
cir = dq.QubitCircuit(3)
# Add a Hadamard gate
cir.h(wires=0)
# Add CNOT gates
cir.cnot(control=0, target=1)
cir.x(wires=2, controls=0)
# Add Ry gates to encode data
cir.rylayer(encode=True)
# Add a layer of CNOT gates in a cyclic way
cir.cnot_ring()
# Add Ry gates with variational parameters
cir.rylayer()
# Set the observable to be measured
cir.observable(wires=[0, 1, 2], basis='xzx')
# Execute the circuit via forward()
cir(data)
# Print the measurement results
print(cir.measure(shots=1000, with_prob=True, wires=[0]))
# Print the expectation values
exp = cir.expectation()
print('Expectation values:')
print(exp)
# Print the gradients
exp.backward()
print('Gradients:')
for param in cir.parameters():
    print(param.grad)
# Draw the circuit
cir.draw()
\end{lstlisting}

\noindent Output:
\begin{verbatim}
{'0': (850, tensor(0.8387,
     grad_fn=<SelectBackward0>)),
 '1': (150, tensor(0.1613,
     grad_fn=<SelectBackward0>))}
Expectation values:
tensor([0.2592], grad_fn=<StackBackward0>)
Gradients:
tensor(0.2225)
tensor(0.6533)
tensor(0.5036)
\end{verbatim}

\begin{figure}[!htbp]
    \centering
    \includegraphics[width=\linewidth]{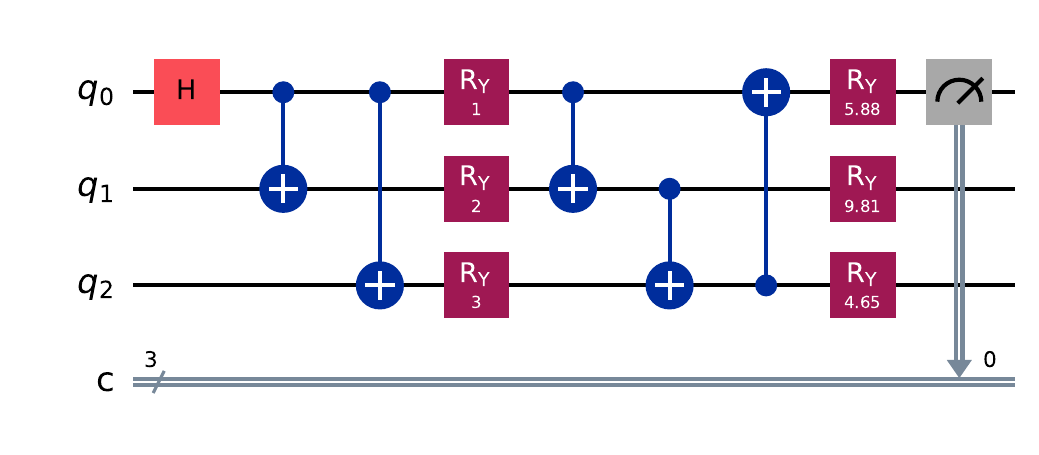}
    \caption{Visualization of the 3-qubit \texttt{QubitCircuit} constructed in Code~\ref{code:qubitcircuit}.}
    \label{fig:qubitcircuit}
\end{figure}

Code~\ref{code:qubitcircuit} presents a representative workflow utilizing \texttt{QubitCircuit}, with the corresponding circuit structure visualized in Fig.~\ref{fig:qubitcircuit}.
As demonstrated, the framework natively supports automatic differentiation through PyTorch, enabling straightforward gradient computation for variational algorithms.
Furthermore, \texttt{QubitCircuit} inherently supports batch processing, allowing it to be seamlessly and efficiently integrated into broader PyTorch-based deep learning pipelines for large-scale quantum-classical simulations.

\subsection{Applications}
\subsubsection{QResNet}
\label{sec:qresnet}
Quantum residual neural networks (QResNets) bring the classical deep learning innovation of residual connections into the quantum computing domain.
Classically, ResNets alleviate the problem of vanishing gradients by introducing shortcut connections~\cite{he2016deep}.
Similarly, in quantum computing, incorporating residual connections serves the dual purpose of enhancing the expressive capacity of quantum neural networks (QNNs)~\cite{wen2024enhancing} and mitigating the barren plateau problem inherent in QML~\cite{kashif2024resqnets}.

Residual connections provide a compact mechanism for enriching the function class represented by data-encoding QNNs.
For a data-encoding unitary $\hat{U}(x)$, the residual construction considered in Ref.~\cite{wen2024enhancing} can be written as
\begin{equation}
    \hat{R}(x)\ket{\phi} = \frac{1}{2}\left(I^{\otimes n} + \hat{U}(x)\right)\ket{\phi}.
\end{equation}
Practically, such a residual operator can be realized via the linear combination of unitaries (LCU) technique, utilizing an ancilla qubit and controlled operations.
The desired residual branch is then obtained by conditioning on a specific ancilla measurement outcome.
Consequently, this mechanism enables the encoded state to coherently superpose the original branch with the unitarily transformed branch.

The significance of this construction becomes transparent from the Fourier perspective of quantum models.
The output of a conventional data-encoding QNN for a measured observable $\hat{O}$ can be expressed as
\begin{equation}
    f(x,\theta) = \sum_{\omega\in\Omega} c_{\omega}(\theta,\hat{O}) e^{i\omega x},
\end{equation}
where the accessible frequency set $\Omega$ is determined by differences between eigenvalues of the encoding generator.
By contrast, the expectation value evaluated with the unnormalized residual operator contains additional interference terms:
\begin{equation}
    \begin{aligned}
        f_R(x,\theta) = \frac{1}{4}\Big(&\mel{\phi}{\hat{O}}{\phi} + \mel{\phi}{\hat{U}^\dagger(x)\hat{O}\hat{U}(x)}{\phi} \\
        &+ 2\mathrm{Re}\mel{\phi}{\hat{O}\hat{U}(x)}{\phi}\Big).
    \end{aligned}
\end{equation}
These cross-terms can introduce frequency components that are absent in the ordinary encoding model, thereby expanding the model's expressivity.

\begin{lstlisting}[caption={Implementation of a traditional QNN and a QResNet for curve fitting.}, label={code:qresnet}]
import deepquantum as dq
from torch import nn

class QNN(nn.Module):
    def __init__(self, residual=False):
        super().__init__()
        self.residual = residual
        self.build_circuit()

    def build_circuit(self):
        if self.residual:
            self.qnn = dq.QubitCircuit(2)
            self.qnn.ry(wires=0)
            self.qnn.u3(wires=1)
            # Encode data
            self.qnn.ry(wires=1, encode=True, controls=0)
            self.qnn.ry(wires=0)
            self.qnn.u3(wires=1)
            self.qnn.observable(wires=1, basis='z')
            self.qnn.observable(wires=[0, 1], basis='zz')
        else:
            self.qnn = dq.QubitCircuit(1)
            self.qnn.u3(wires=0)
            # Encode data
            self.qnn.ry(wires=0, encode=True)
            self.qnn.u3(wires=0)
            self.qnn.observable(wires=0, basis='z')

    def forward(self, x):
        self.qnn(data=x)
        exp = self.qnn.expectation()
        if self.residual:
            exp = (exp[:, [0]] + exp[:, [1]]) / 2
        return exp
\end{lstlisting}

To demonstrate this structural advantage in a minimal setting, we compare a standard one-qubit QNN with a residual QNN on a one-dimensional curve-fitting task whose target contains an additional $\omega=0.5$ component inaccessible to the baseline encoding.
The DeepQuantum implementation is shown in Code~\ref{code:qresnet}.
Here, the controlled encoding branch realizes the residual construction introduced above, while the average of the $Z_1$ and $Z_0Z_1$ readouts implements the corresponding unnormalized ancilla-conditioned expectation value.

\begin{figure}[!htbp]
    \centering
    \includegraphics[width=\linewidth]{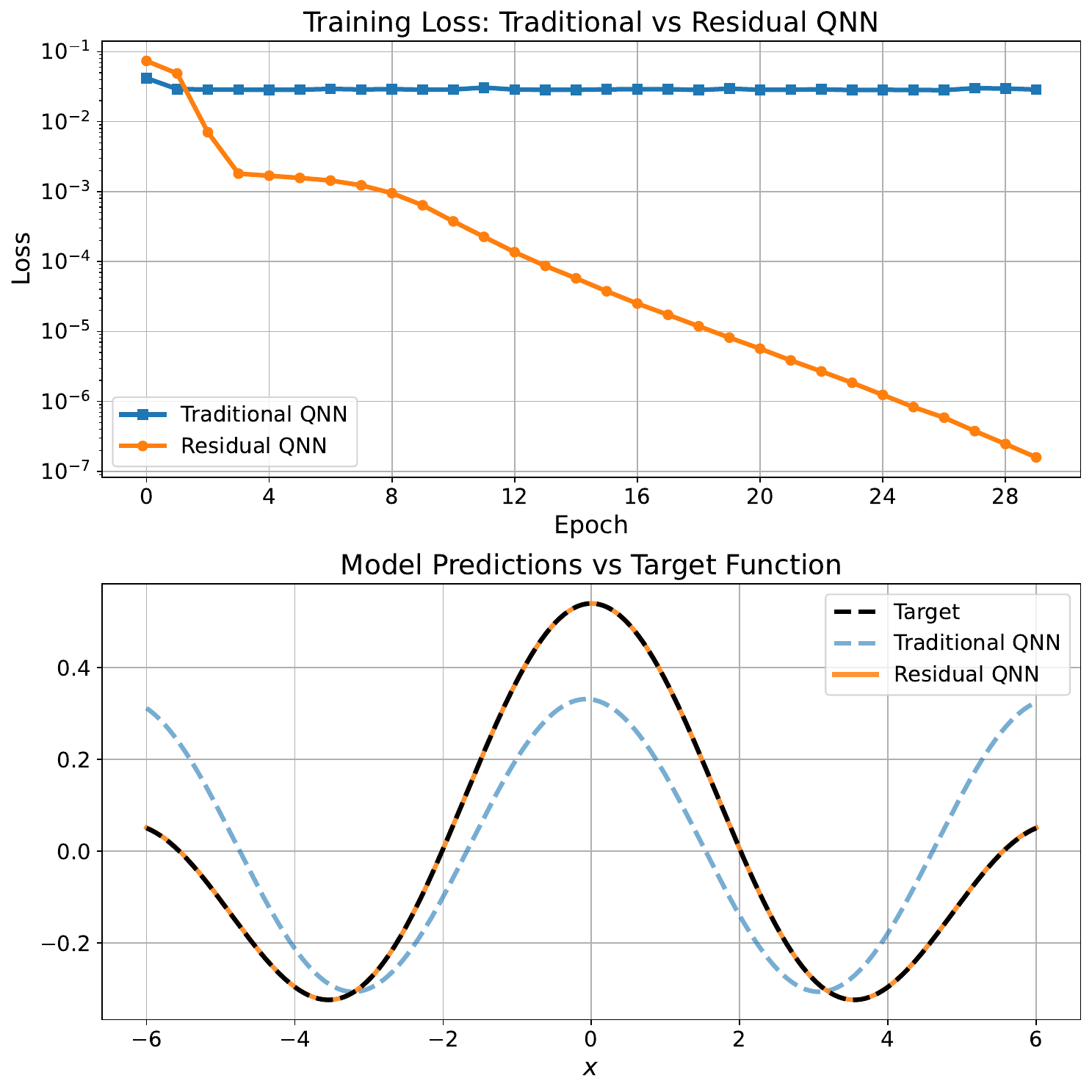}
    \caption{Performance comparison between the traditional QNN and the residual QNN on a curve-fitting task.
    (a) Training loss curves plotted on a logarithmic scale.
    The residual QNN maintains an exponential decay, while the traditional QNN stagnates early.
    (b) Comparison of the final model predictions against the target function.
    The residual QNN near-perfectly captures the target curve, whereas the traditional QNN suffers from limited expressivity and fails to reconstruct the full amplitude.}
    \label{fig:qresnet_train}
\end{figure}

The resulting training curves and fitted functions are reported in Fig.~\ref{fig:qresnet_train}.
As shown in Fig.~\ref{fig:qresnet_train}(a), the residual model continues to reduce the fitting error by several orders of magnitude, whereas the baseline QNN quickly saturates.
Fig.~\ref{fig:qresnet_train}(b) further shows that the residual model closely reproduces the target profile in this controlled example, consistent with the broader accessible frequency spectrum enabled by residual encoding.

\subsubsection{QCNN}
\label{sec:qcnn}
The quantum convolutional neural network (QCNN) is a hybrid quantum algorithm inspired by classical convolutional neural network architectures.
It consists of interleaved convolutional layers, which apply local parameterized unitary operations, and pooling layers, which perform measurements on a subset of qubits to reduce the system's dimensionality.
This architecture allows for efficient training and scalability~\cite{Cong2019}.
While initially proposed for quantum state classification~\cite{Cong2019}, QCNNs have also been successfully applied to classical tasks~\cite{Hur2022, Minu2023, Chen2023}.

\begin{lstlisting}[caption={Implementation of a QCNN for MNIST binary classification.}, label={code:qcnn_demo}]
import deepquantum as dq
import torch
import torch.nn.functional as F
from torch import nn

class QuantumCNN(nn.Module):
    def __init__(self, nqubit=12):
        super().__init__()
        self.nqubit = nqubit
        self.qc = dq.QubitCircuit(nqubit)
        self.build_circuit()
        self.qc.observable(wires=0, basis='z')

    def build_circuit(self):
        # Encode features hierarchically between variational layers
        self.qc.rylayer(encode=True)
        self.add_conv_layer(12)
        self.add_pool_layer(12)
        self.qc.rylayer(wires=list(range(6)), encode=True)
        self.add_conv_layer(6)
        self.add_pool_layer(6)
        self.qc.rylayer(wires=list(range(3)), encode=True)
        self.add_conv_layer_3qubits()
        self.add_pool_layer_3to1()

    # The following layers are variational
    def add_conv_layer(self, nqubit):
        for i in range(0, nqubit - 1, 2):
            self.qc.rxx([i, i + 1])
            self.qc.ryy([i, i + 1])

    def add_conv_layer_3qubits(self):
        for i in range(2):
            self.qc.rxx([i, i + 1])
            self.qc.ryy([i, i + 1])

    def add_pool_layer(self, nqubit):
        for i in range(nqubit // 2):
            control = 2 * i + 1
            target = 2 * i
            self.qc.cry(control, target)

    def add_pool_layer_3to1(self):
        self.qc.cry(1, 0)
        self.qc.cry(2, 0)

    # Batch forward pass
    def forward(self, x):
        batch_size = x.shape[0]
        x = x.view(batch_size, 1, 28, 28)
        # Pool to 3x7 = 21 features to match the 12+6+3 encoding gates
        features = F.adaptive_avg_pool2d(x, (3, 7)).view(batch_size, -1)
        angle = torch.pi * torch.tanh(features)
        self.qc(data=angle)
        # Inputs are encoded in batches
        expectation = self.qc.expectation()
        output = expectation * 3.0
        return output.squeeze()
\end{lstlisting}

We use DeepQuantum to implement a QCNN for a binary image classification task based on the MNIST dataset.
Specifically, our aim is to distinguish between handwritten digit images labeled as 0 and 1.
To achieve this, the input image data is downsampled and encoded into a 12-qubit system.
The classification output is determined by the sign of the expectation value of the Pauli-$Z$ observable, which is measured on the designated readout qubit at the end of the circuit.
The core implementation of the QCNN, which follows the purely quantum model architecture proposed in Ref.~\cite{Hur2022}, is detailed in Code~\ref{code:qcnn_demo}.
Notably, the $28 \times 28$ input image is adaptively pooled into 21 continuous features, which are hierarchically uploaded into the quantum circuit via $R_y$ layers acting on 12, 6, and 3 qubits progressively.

\begin{figure}[!htbp]
    \centering
    \includegraphics[width=\linewidth]{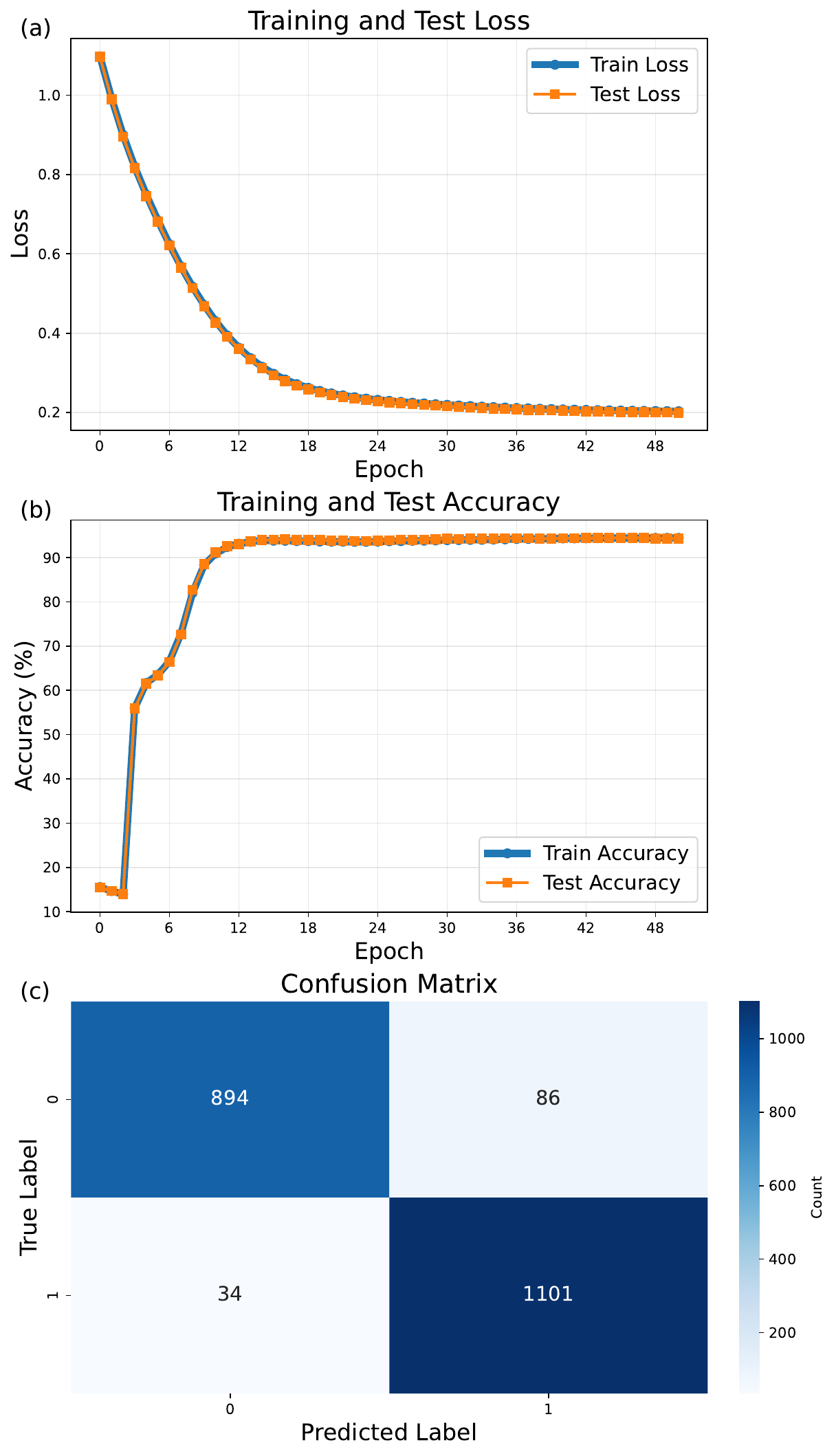}
    \caption{Performance of the QCNN on the MNIST binary classification task.
    (a) Training and test loss convergence over 50 epochs.
    (b) Corresponding training and test accuracy, reaching over 94\%.
    (c) Confusion matrix evaluated on 2115 test samples, detailing the distribution of correct and incorrect predictions.}
    \label{fig:qcnn}
\end{figure}

The learning dynamics and performance metrics of our QCNN model are summarized in Fig.~\ref{fig:qcnn}.
As depicted in Fig.~\ref{fig:qcnn}(a) and (b), both the loss function and accuracy metrics converge smoothly over 50 training epochs.
The model achieves a final training accuracy of 94.46\%, with the test curve closely tracking the training curve, indicating stable learning without significant overfitting.
Evaluated on a test set of 2115 unseen samples, the trained model achieves a robust test accuracy of 94.33\%.
The corresponding confusion matrix, presented in Fig.~\ref{fig:qcnn}(c), further breaks down the model's predictive performance, demonstrating its strong capability in discriminating between the two classes with low false positive and false negative rates.

\section{Photonic Quantum Computing}
\label{sec:qumode}
\subsection{Background}
Photonic quantum computing, which utilizes the properties of photons such as polarization and coherence, has emerged as a leading candidate for building scalable and efficient quantum processors~\cite{kok2007linear}.
Since photons interact weakly with their environment, they allow for robust quantum states that are highly resilient to decoherence, making them ideal for quantum information processing~\cite{o2007optical}.
Furthermore, photons enable the transmission of quantum information over long distances, which is crucial for building quantum networks and integrating quantum computing into real-world applications~\cite{gisin2007quantum}.
In photonic quantum computing and communication, DV and CV systems constitute the two primary paradigms.

In DV architectures, quantum information is encoded in discrete degrees of freedom of light, typically represented by Fock states $\ket{n}$.
These states are eigenstates of the photon-number operator $\hat{N}$:
\begin{equation}
    \hat{N}\ket{n} = n\ket{n}.
\end{equation}
DeepQuantum provides two distinct Fock backends for simulating DV systems: one based on Fock basis states, and the other on Fock state tensors.

The first approach models linear optical networks via matrix permanents.
Consider an initial $N$-mode Fock state $\ket{\psi_{\mathrm{in}}} = \ket{s_1, \dots, s_N}$ incident on an $N$-mode linear interferometer described by a unitary matrix $U$.
The transition probability to an output Fock state $\ket{\psi_{\mathrm{out}}} = \ket{t_1, \dots, t_N}$ is governed by the permanent of a submatrix $U_{T,S}$.
Specifically, $U_{T,S}$ is constructed by taking $s_i$ copies of the $i$-th column and $t_j$ copies of the $j$-th row of $U$.
The probability is given by
\begin{equation}
    \abs{\mel{\psi_{\mathrm{out}}}{\hat{\mathcal{U}}}{\psi_{\mathrm{in}}}}^2 = \frac{\abs{\mathrm{Per}(U_{T,S})}^2}{\prod_{i=1}^N s_i! \prod_{j=1}^N t_j!},
\end{equation}
where $\hat{\mathcal{U}}$ represents the unitary evolution operator acting on the many-body Fock space, induced by the linear transformation $U$.

The second approach addresses the simulation of systems involving arbitrary superpositions of Fock states.
It employs a tensor representation for the system's evolution, serving as a direct generalization of multi-qubit simulation methods.
For an $N$-mode system with a local Fock-space truncation dimension $d$, a pure state is represented as an order-$N$ tensor with shape $(d,\dots,d)$, while a density matrix is represented as an order-$2N$ tensor.
Correspondingly, a $k$-mode photonic quantum gate is expressed as an order-$2k$ tensor.
The system's time evolution is executed by applying the gate tensor to the state tensor via tensor contraction over the indices associated with the target modes.

In contrast, CV architectures utilize the continuous degrees of freedom of the electromagnetic field.
These are defined via the canonical annihilation and creation operators, $\hat{a}$ and $\hat{a}^\dagger$, and are commonly visualized in phase space using the position and momentum quadrature operators $\hat{x}$ and $\hat{p}$~\cite{braunstein2005quantum}:
\begin{equation}
    \begin{aligned}
        \hat{x} &\equiv \frac{\sqrt{\hbar}}{2\kappa}(\hat{a}+\hat{a}^{\dagger}), \\
        \hat{p} &\equiv \frac{\sqrt{\hbar}}{2i\kappa}(\hat{a}-\hat{a}^{\dagger}).
    \end{aligned}
\end{equation}
Here the quadratures $\hat{x}$ and $\hat{p}$ possess the physical dimension of $\sqrt{\hbar}$.
In DeepQuantum, we set $\hbar=2$ and $\kappa=\frac{1}{\sqrt{2}}$ by default.
For the sake of generality, subsequent phase-space expressions explicitly retain $\hbar$ under this $\kappa$ convention, thereby yielding the standard commutation relation $\comm{\hat{x}}{\hat{p}} = i\hbar$.

The Wigner function provides a powerful phase-space representation of quantum states, serving as a bridge between quantum mechanics and classical statistical physics.
For a single-mode quantum state described by a density matrix $\rho$, its Wigner function is defined as
\begin{equation}
    W(x,p) = \frac{1}{\pi\hbar} \int_{-\infty}^{\infty} \mel{x - y}{\rho}{x + y} e^{2ipy / \hbar} \dd{y},
\end{equation}
where $x$ and $p$ denote the field quadratures.
As a real-valued quasi-probability distribution, $W(x,p)$ can exhibit negative values, which are strong signatures of non-classicality in CV systems.
Furthermore, the expectation value of any observable $\hat{O}$ can be computed via phase-space integration:
\begin{equation}
    \expval{\hat{O}} = \iint W(x,p)O_W(x,p) \dd{x} \dd{p},
\end{equation}
where the corresponding Weyl symbol $O_W(x,p)$ is given by
\begin{equation}
    O_W(x,p) = 2 \int_{-\infty}^{\infty} \mel{x + y}{\hat{O}}{x - y} e^{-2ipy / \hbar} \dd{y}.
\end{equation}

The single-mode Wigner function generalizes naturally to multimode systems.
For an $m$-mode quantum state $\rho$, the Wigner function is the symplectic Fourier transform of its characteristic function $\chi_\rho(\eta) = \Tr[\rho \hat{D}(\eta)]$:
\begin{equation}
    W_\rho(\xi) = \frac{1}{(2\pi\hbar)^{2m}} \int_{\mathbb{R}^{2m}} e^{-i \xi^T \Omega \eta / \hbar} \chi_\rho(\eta) \dd^{2m}{\eta},
\end{equation}
where $\xi = (x_1, \dots, x_m, p_1, \dots, p_m)^T$ is the phase-space coordinate vector.
Here, $\hat{D}(\eta)=\exp(i\hat{R}^T\Omega\eta / \hbar)$ is the Weyl displacement operator, $\hat{R} = (\hat{x}_1, \dots, \hat{x}_m, \hat{p}_1, \dots, \hat{p}_m)^T$ is the vector of quadrature operators, and the symplectic form is defined as
\begin{equation}
    \Omega =
    \begin{pmatrix}
        0 & I_m \\
        -I_m & 0
    \end{pmatrix}.
\end{equation}
This formalism provides a complete description of quantum states in phase space and is particularly efficient for simulating Gaussian states and operations~\cite{bartlett2002efficient}.

Gaussian states, including coherent and squeezed states, are ubiquitous in CV quantum information protocols such as quantum key distribution, teleportation, and entanglement distribution~\cite{braunstein2005quantum, weedbrook2012gaussian, ferraro2005gaussian}.
A Gaussian state is defined as any quantum state whose Wigner function is a Gaussian distribution:
\begin{equation}
    W_\rho(\xi) = \frac{1}{(2\pi)^m \sqrt{\det V}} \exp\left[-\frac{1}{2}(\xi - R)^T V^{-1} (\xi - R)\right],
\end{equation}
where $R = \expval*{\hat{R}}$ is the mean displacement vector, and $V$ is the covariance matrix with elements $V_{ij} = \frac{1}{2}\expval*{\hat{R}_i\hat{R}_j + \hat{R}_j\hat{R}_i} - R_i R_j$.
Within DeepQuantum's Gaussian backend, any Gaussian state is efficiently characterized solely by the pair $(V, R)$.

Many fundamental quantum optical components, such as phase shifters and squeezers, implement Gaussian transformations.
A transformation is considered Gaussian if it is generated by a unitary operator of the form $\hat{U}=\exp(-i\hat{H}/\hbar)$, where the Hamiltonian $\hat{H}$ is at most quadratic in the quadrature operators.
In the Heisenberg picture, the evolution of the quadrature vector under such a unitary corresponds to an affine symplectic transformation in phase space:
\begin{equation}
    \hat{U}^\dagger \hat{R} \hat{U} = S \hat{R} + D,
\end{equation}
where $S$ is a symplectic matrix preserving the symplectic form $S\Omega S^T=\Omega$, and $D$ is a real displacement vector.

However, achieving universal quantum computation in the CV regime requires the injection of non-Gaussian resources, such as photon addition, photon subtraction, or the preparation of specific non-Gaussian ancilla states~\cite{lloyd1999quantum, bartlett2002universal, ourjoumtsev2007generation}.
To address this, DeepQuantum utilizes the Bosonic backend to simulate non-Gaussian states $\rho$ whose Wigner functions can be decomposed as a linear combination of Gaussian functions~\cite{bourassa2021fast}:
\begin{equation}
    W_\rho(\xi) = \sum_i w_i G^{(i)}(\xi),
\end{equation}
where each $G^{(i)}(\xi)$ is a generally complex normalized Gaussian function characterized by its covariance matrix $V^{(i)}$ and mean vector $R^{(i)}$, and the complex weights $w_i$ satisfy the normalization condition $\sum_i w_i = 1$.
Consequently, such a non-Gaussian state is parameterized by the set of tuples $\{(V^{(i)}, R^{(i)}, w_i)\}$.

Two typical non-Gaussian states are the cat state and the GKP state.
The cat state is formed by the quantum superposition of two opposite coherent states:
\begin{equation}
    \ket{\mathrm{Cat}} = \frac{1}{\mathcal{N}_\phi} (\ket{\alpha} + e^{i\phi}\ket{-\alpha}),
\end{equation}
where $\ket{\alpha} = \hat{D}(\alpha)\ket{0}$, and $\mathcal{N}_\phi = \sqrt{2+2e^{-2\abs{\alpha}^2}\cos\phi}$ is the normalization factor.
Setting $\phi=0$ yields the even cat state, while $\phi=\pi$ produces the odd cat state.
The GKP state~\cite{gottesman2001encoding} is a specialized quantum error-correcting code that encodes logical qubits into grid-like superpositions in phase space.
An arbitrary logical GKP state is defined as
\begin{equation}
    \ket{\psi}_{\mathrm{GKP}} = \cos\frac{\theta}{2}\ket{0}_{\mathrm{GKP}} + e^{-i\phi}\sin\frac{\theta}{2}\ket{1}_{\mathrm{GKP}}.
\end{equation}
In the ideal limit, the Wigner functions of the logical basis states $\ket{0}_{\mathrm{GKP}}$ and $\ket{1}_{\mathrm{GKP}}$ manifest as a discrete grid of Dirac $\delta$-functions~\cite{garcia2020bloch}:
\begin{equation}
    W_{\mathrm{GKP}}^0(x,p) \propto \sum_{s,t=-\infty}^{\infty} (-1)^{st} \delta\pqty{p-\frac{s\sqrt{\pi\hbar}}{2}} \delta\pqty{x-t\sqrt{\pi\hbar}},
\end{equation}
\begin{equation}
    W_{\mathrm{GKP}}^1(x,p) \propto \sum_{s,t=-\infty}^\infty (-1)^{st} \delta\pqty{p-\frac{s\sqrt{\pi\hbar}}{2}} \delta\pqty{x-(t+1)\sqrt{\pi\hbar}}.
\end{equation}

\subsection{API Overview}
\begin{table*}[!htbp]
\centering
\caption{Comparison of supported features across the three photonic backends in DeepQuantum.}
\label{tab:qumode_backends}
\tabcolsep=0.4cm
\begin{tabular}{@{}l c c c c@{}}
\toprule
  \textbf{Backend}& \multicolumn{2}{c}{\textbf{Fock}} & \textbf{Gaussian} & \textbf{Bosonic} \\
\cmidrule(lr){2-3}
  & Basis state & State tensor & & \\
\midrule
\textbf{State representation} &
  Fock basis &
  \makecell{State vector \\ Density matrix} &
  $(V, R)$ &
  $\{(V^{(i)}, R^{(i)}, w_i)\}$ \\
\addlinespace
\textbf{Detector} &
  PNRD &
  \makecell{PNRD \\ Homodyne} &
  \makecell{PNRD \\ Threshold \\ Homodyne} &
  Homodyne \\
\addlinespace
\textbf{Channel} &
  Photon loss &
  Photon loss &
  Photon loss &
  Photon loss \\
\bottomrule
\end{tabular}
\end{table*}

In DeepQuantum, the central class \texttt{QumodeCircuit} is designed for simulating photonic quantum circuits.
The general simulation workflow consists of four sequential stages: initializing the circuit with quantum states, constructing the circuit by applying gates, evolving the system, and finally, sampling the results through measurement.
Furthermore, to model realistic experimental conditions, noise channels such as photon loss can be integrated directly into the circuit.
These processes are supported across three distinct backends: the Fock, Gaussian and Bosonic backends.
A comprehensive comparison of their supported features is summarized in Table~\ref{tab:qumode_backends}.

To facilitate this workflow, the DeepQuantum API is organized into four components:
\begin{itemize}
    \item \textbf{Quantum state.}
    Users can initialize quantum states utilizing various representations, including DV Fock states, CV Gaussian states, and non-Gaussian Bosonic states.

    \item \textbf{Quantum operation.}
    A comprehensive library of single- and two-mode gates, as well as noise channels, is provided.
    Typical operations include phase shifters, squeezers, beam splitters, Mach-Zehnder interferometers (MZIs), and photon loss channels.

    \item \textbf{Quantum circuit.}
    The \texttt{QumodeCircuit} class provides methods to assemble gate sequences, execute forward evolutions, and visualize circuit structures.
    Forward execution (invoked via \texttt{QumodeCircuit.forward()}) produces the output quantum state, with the underlying data structure depending on the selected backend.

    \item \textbf{Measurement.}
    The measurement module supports photon-number-resolving detectors (PNRD), threshold detectors, and homodyne measurements.
    Depending on the chosen backend, different sampling methods are available, enabling the extraction of statistical outcomes, exact probabilities, and quadrature distributions.
\end{itemize}

\begin{lstlisting}[caption={An illustrative example of \texttt{QumodeCircuit} using the Gaussian backend.}, label={code:qumodecircuit}]
import deepquantum as dq
import torch

# Define trainable parameters
r0 = torch.nn.Parameter(torch.tensor(0.5))
r1 = torch.nn.Parameter(torch.tensor(1.0))
theta = torch.nn.Parameter(torch.tensor(1.0))
# Initialize a 2-mode circuit; the photon-number cutoff of 5 is used for sampling
cir = dq.QumodeCircuit(nmode=2, init_state='vac', cutoff=5, backend='gaussian')
# Add squeezing gates
cir.s(0, r=r0)
cir.s(1, r=r1)
# Add a beam splitter
cir.bs([0, 1], inputs=[theta, 1.0])
# Forward process
cir()
# Print measurement results with different detectors
print(cir.measure(shots=100, detector='threshold'))
print(cir.measure(shots=100, detector='pnrd'))
# Set homodyne measurement on mode 0 with phase pi/4
cir.homodyne(wires=[0], phi=torch.pi/4)
cir()
# Print homodyne measurement results
sample = cir.measure_homodyne(shots=1024)
print(sample)
# Forward execution and return probability distribution with gradients
cir(is_prob=True)
# Draw the circuit
cir.draw()
\end{lstlisting}

\noindent Output:
\begin{verbatim}
Using chain-rule method to sample the final states!
{|00>: 51, |11>: 34, |01>: 8, |10>: 7}
Using chain-rule method to sample the final states!
{|02>: 17, |13>: 15, |04>: 3, |22>: 1, |40>: 1,
 |44>: 2, |20>: 1, |00>: 55, |42>: 5}

tensor([1.7850,  2.6191, -0.2665,  ...,
        3.1729, -0.5978, -0.1189])

{|00>: tensor([0.5747],
     grad_fn=<UnbindBackward0>),
 |02>: tensor([0.0276],
     grad_fn=<UnbindBackward0>),
 |11>: tensor([0.1291],
     grad_fn=<UnbindBackward0>),
 ......,
 |41>: tensor([0.],
     grad_fn=<UnbindBackward0>),
 |34>: tensor([0.],
     grad_fn=<UnbindBackward0>),
 |43>: tensor([0.],
     grad_fn=<UnbindBackward0>)}
\end{verbatim}

\begin{figure}[!htbp]
    \centering
    \includegraphics[width=\linewidth]{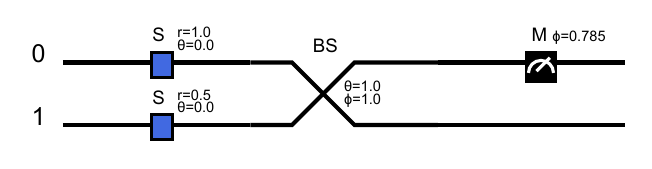}
    \caption{Visualization of the 2-mode \texttt{QumodeCircuit} constructed in Code~\ref{code:qumodecircuit}.}
    \label{fig:qumodecircuit}
\end{figure}

Together, these components form a coherent and end-to-end differentiable framework that allows users to construct photonic quantum algorithms in a flexible and efficient manner.
To demonstrate how these components interact in practice, Code~\ref{code:qumodecircuit} presents a complete workflow using the Gaussian backend, with the corresponding circuit structure visualized in Fig.~\ref{fig:qumodecircuit}.
Notably, similar to the \texttt{QubitCircuit} module, \texttt{QumodeCircuit} natively supports automatic differentiation through PyTorch.
This feature enables gradient-based optimization directly from differentiable probability outputs, as indicated by the \texttt{grad\_fn} attributes in the output block.

\begin{lstlisting}[caption={Initialization of non-Gaussian states using the Bosonic backend.}, label={code:bosonic_init}]
cir = dq.QumodeCircuit(nmode=2, init_state='vac', backend='bosonic')
cir.cat(wires=0, r=1.0)
cir.gkp(wires=1, theta=0.0, amp_cutoff=0.1)
state = cir()
for s in state:
    print(s.shape)
\end{lstlisting}

\noindent Output:
\begin{verbatim}
torch.Size([1, 1, 4, 4])
torch.Size([1, 356, 4, 1])
torch.Size([1, 356])
\end{verbatim}

Furthermore, the Bosonic backend seamlessly supports the initialization and evolution of highly non-classical states, such as cat states and GKP states, as demonstrated in Code~\ref{code:bosonic_init}.
The multi-dimensional tensor shapes in the output directly correspond to the internal representation of the Bosonic backend.
Here, the system is parameterized by its covariance matrices, mean vectors, and complex weights, reflecting the underlying Gaussian decomposition.

\subsection{Time-Domain Multiplexing}
\begin{figure}[!htbp]
    \centering
    \includegraphics[width=\linewidth]{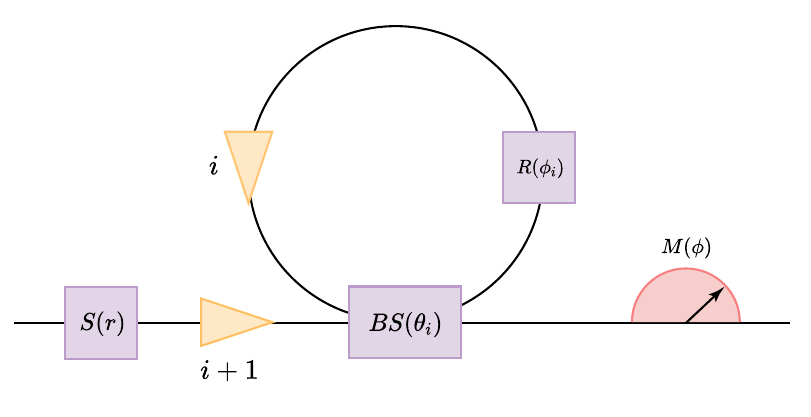}
    \caption{Canonical setup of a TDM photonic quantum circuit.}
    \label{fig:tdm_setup}
\end{figure}

Photons, serving as flying qubits, can be precisely delayed in optical media---a critical capability that underpins TDM.
Leveraging this technique, time-domain-multiplexed photonic quantum circuits can generate large-scale, highly entangled CV quantum states, such as cluster states encompassing more than 10,000 entangled modes~\cite{yokoyama2013ultra, alexander2016flexible, asavanant2019generation}.
A canonical time-domain photonic setup is illustrated in Fig.~\ref{fig:tdm_setup}.
In this architecture, the incoming $(i+1)$-th pulse is initially squeezed by $S(r)$ and subsequently interferes with the preceding $i$-th pulse (which emerges from the delay loop) at a tunable beam splitter $BS(\theta_i)$.
Following this interference, the output is split into two distinct paths.
One portion is directed back into the delay loop, passing through a phase shifter $R(\phi_i)$ to interfere with the next incoming pulse.
The remaining portion exits the loop and proceeds to a homodyne measurement $M(\phi)$.
Naively simulating this sequential temporal process of $N$ steps requires mapping it to $(N+1)$ spatial modes.
For a large number of pulses $N$, this explicit spatial mapping becomes computationally prohibitive, demanding excessive memory and execution time.

To address this scalability bottleneck, DeepQuantum introduces a natively reconfigurable delay loop abstraction.
The delay loop, serving as the core structural element for TDM, is coupled to the main optical waveguide via a tunable beam splitter, thereby intrinsically introducing a precise time delay.
Critically, as depicted in Fig.~\ref{fig:tdm_setup}, a phase shifter is embedded within the loop to dynamically manipulate the delayed quantum state prior to its subsequent interference.

\begin{lstlisting}[caption={An illustrative example of setting up a TDM photonic quantum circuit.}, label={code:tdm_circuit}]
import deepquantum as dq
import torch

cir = dq.QumodeCircuit(nmode=1, init_state='vac', cutoff=3, backend='gaussian')
cir.s(0)
cir.delay(0, ntau=1, inputs=[torch.pi/4, 0])
cir.r(0)
cir.homodyne(0)
cir.draw()
\end{lstlisting}

\begin{figure}[!htbp]
    \centering
    \includegraphics[width=\linewidth]{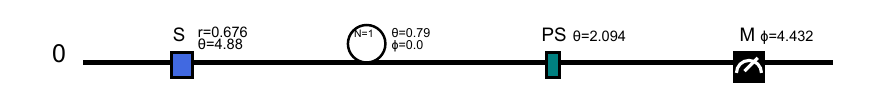}
    \caption{The TDM photonic quantum circuit generated in Code~\ref{code:tdm_circuit}.}
    \label{fig:tdm_circuit}
\end{figure}

In DeepQuantum, a delay loop can be seamlessly instantiated within a standard \texttt{QumodeCircuit} utilizing the \texttt{delay()} method, as demonstrated in Code~\ref{code:tdm_circuit}.
The generated compact circuit structure is visualized in Fig.~\ref{fig:tdm_circuit}.
The \texttt{delay()} method is configured by several arguments, most notably \texttt{ntau}, which defines the delay length (i.e., the number of concurrent modes stored in the loop), and \texttt{inputs}, which parameterizes the loop's internal operations, such as the beam splitter's transmissivity and the internal phase shift.

\begin{lstlisting}[caption={Visualizing the unrolled equivalent spatial circuit.}, label={code:tdm_unroll}]
cir.draw(unroll=True)
\end{lstlisting}

\begin{figure}[!htbp]
    \centering
    \includegraphics[width=\linewidth]{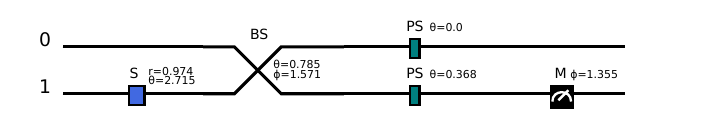}
    \caption{Equivalent unrolled spatial circuit corresponding to a single-step evolution of the TDM circuit from Fig.~\ref{fig:tdm_circuit}.}
    \label{fig:tdm_unroll}
\end{figure}

Because the standard \texttt{QumodeCircuit} is designed to model static, single-step evolutions, TDM simulations at this level are realized by unrolling the delay loops into an equivalent single-step spatial circuit.
This approach allows users to explicitly match the initial states and track the corresponding output modes.
Furthermore, by passing the \texttt{unroll=True} argument to the \texttt{draw()} method, users can directly visualize this equivalent spatial mapping, as shown in Code~\ref{code:tdm_unroll} and Fig.~\ref{fig:tdm_unroll}.

\begin{lstlisting}[caption={Executing \texttt{QumodeCircuitTDM} and compiling the full temporal evolution into a global spatial circuit.}, label={code:tdm_global}]
cir = dq.QumodeCircuitTDM(nmode=1, init_state='vac', cutoff=3)
cir.s(0)
cir.delay(0, ntau=1, inputs=[torch.pi/4, 0])
cir.r(0)
cir.homodyne(0)
# Forward execution for 2 sequential steps
cir(nstep=2)
# Get the homodyne measurement samples for the 2 steps
print(cir.samples)
# Compile and draw the global spatial circuit
cir_global = cir.global_circuit(nstep=2)
cir_global.draw(unroll=True)
\end{lstlisting}

\noindent Output:
\begin{verbatim}
tensor([-3.1730, 0.9339])
\end{verbatim}

\begin{figure}[!htbp]
    \centering
    \includegraphics[width=\linewidth]{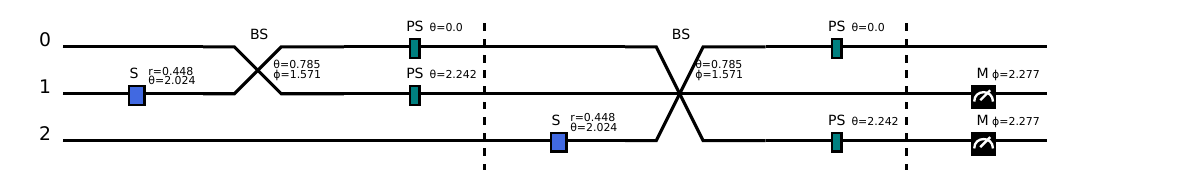}
    \caption{Equivalent global spatial circuit for the full multi-step TDM evolution, mapped to \texttt{nstep+1} spatial modes.}
    \label{fig:tdm_global}
\end{figure}

However, to simulate the complete, multi-step temporal dynamics, users should employ the specialized \texttt{QumodeCircuitTDM} class.
This class manages the full temporal evolution and requires that all spatial modes are measured at each time step, as demonstrated in Code~\ref{code:tdm_global}.
To provide deeper physical insight into this temporal evolution, users can invoke the \texttt{global\_circuit()} method, which mathematically compiles the multi-step temporal process into an equivalent global spatial \texttt{QumodeCircuit}.
As illustrated in Fig.~\ref{fig:tdm_global}, for an \texttt{nstep} of 2, the TDM circuit is correctly unrolled into an expansive spatial circuit spanning \texttt{nstep+1} (i.e., 3) spatial modes.

\subsection{Applications}
\label{sec:photonic_apps}
\subsubsection{CNOT Based on Dual-Rail Encoding}
The controlled-NOT ($\mathrm{CNOT}$) gate is a fundamental entangling operation in quantum computing.
It plays a pivotal role in universal quantum computation, enabling the generation of entangled states, the implementation of quantum error correction, and the construction of complex logical circuits.
Combined with single-qubit rotations, the $\mathrm{CNOT}$ gate forms a universal gate set, making it a canonical entangling primitive in universal gate decompositions~\cite{barenco1995elementary, nielsen2010quantum}.

However, implementing a deterministic $\mathrm{CNOT}$ gate on photonic platforms remains a significant challenge.
In contrast to matter-based qubits (e.g., superconducting circuits or trapped ions), photons do not directly interact with each other within linear optical media~\cite{pittman2001probabilistic}.
To overcome this fundamental limitation, the Knill-Laflamme-Milburn (KLM) protocol was proposed~\cite{knill2001scheme}.
It introduces an elegant framework to induce an effective non-linear interaction between photons by leveraging linear optical elements, ancillary photons, and projective measurements.
While the full scalable KLM protocol achieves near-deterministic operations via resource-intensive quantum teleportation and feed-forward mechanisms, foundational and near-term implementations typically rely on more compact, probabilistic schemes that utilize post-selection~\cite{ralph2002linear, o2003demonstration}.

\begin{figure}[!htbp]
    \centering
    \includegraphics[width=\linewidth]{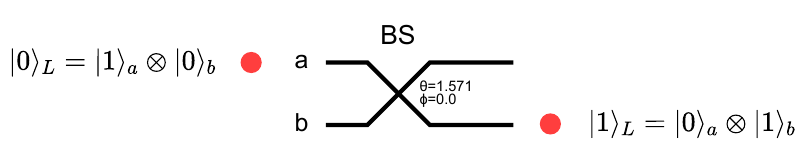}
    \caption{Definition of the dual-rail encoding scheme for photonic qubits.}
    \label{fig:dual_rail}
\end{figure}

To demonstrate this within DeepQuantum, the dual-rail encoding technique~\cite{ralph2002linear} is employed to implement the probabilistic $\mathrm{CNOT}$ gate using the DV framework.
Specifically, each logical qubit is represented by a single photon superposed across two distinct spatial modes.
The logical states are thus determined by the photon's occupancy within these two spatial modes $a$ and $b$, as illustrated in Fig.~\ref{fig:dual_rail}:
\begin{equation}
    \begin{aligned}
        \ket{0}_{L} &\equiv \ket{1}_{a} \otimes \ket{0}_{b}, \\
        \ket{1}_{L} &\equiv \ket{0}_{a} \otimes \ket{1}_{b}.
    \end{aligned}
\end{equation}

The optical implementation of the $\mathrm{CNOT}$ gate relies fundamentally on a probabilistic controlled-Z ($\mathrm{CZ}$) gate.
Mathematically, the $\mathrm{CNOT}$ gate can be constructed from the $\mathrm{CZ}$ gate by applying a Hadamard gate before and after the target qubit~\cite{o2003demonstration}:
\begin{equation}
    \mathrm{CNOT} = (I \otimes H) \cdot \mathrm{CZ} \cdot (I \otimes H).
\end{equation}

\begin{lstlisting}[caption={Implementation of a probabilistic photonic $\mathrm{CNOT}$ gate.}, label={code:cnot}]
import deepquantum as dq
import torch

# The first and last modes are ancilla modes
init_state = [0, 1, 0, 1, 0, 0]
cir = dq.QumodeCircuit(6, init_state)
theta = torch.arccos(torch.tensor(1 / 3**0.5)) * 2
# Hadamard gate
cir.h([3, 4])
# CZ gate
cir.ps(1, torch.pi)
cir.bs_h([0, 1], theta)
cir.ps(0, torch.pi)
cir.ps(3, torch.pi)
cir.bs_h([2, 3], theta)
cir.ps(2, torch.pi)
cir.bs_h([4, 5], theta)
# Hadamard gate
cir.h([3, 4])
print('|00>->|00> amplitude:')
print(
    cir.get_amplitude(
        final_state=[0, 1, 0, 1, 0, 0],
        init_state=[0, 1, 0, 1, 0, 0],
        )
    )
print('|01>->|01> amplitude:')
print(
    cir.get_amplitude(
        final_state=[0, 1, 0, 0, 1, 0],
        init_state=[0, 1, 0, 0, 1, 0],
        )
    )
print('|10>->|11> amplitude:')
print(
    cir.get_amplitude(
        final_state=[0, 0, 1, 0, 1, 0],
        init_state=[0, 0, 1, 1, 0, 0],
        )
    )
print('|11>->|10> amplitude:')
print(
    cir.get_amplitude(
        final_state=[0, 0, 1, 1, 0, 0],
        init_state=[0, 0, 1, 0, 1, 0],
        )
    )
cir.draw()
\end{lstlisting}

\noindent Output:
\begin{verbatim}
|00>->|00> amplitude:
tensor([0.3333+4.3711e-08j])
|01>->|01> amplitude:
tensor([0.3333+4.3711e-08j])
|10>->|11> amplitude:
tensor([0.3333+4.3711e-08j])
|11>->|10> amplitude:
tensor([0.3333+4.3711e-08j])
\end{verbatim}

\begin{figure}[!htbp]
    \centering
    \includegraphics[width=\linewidth]{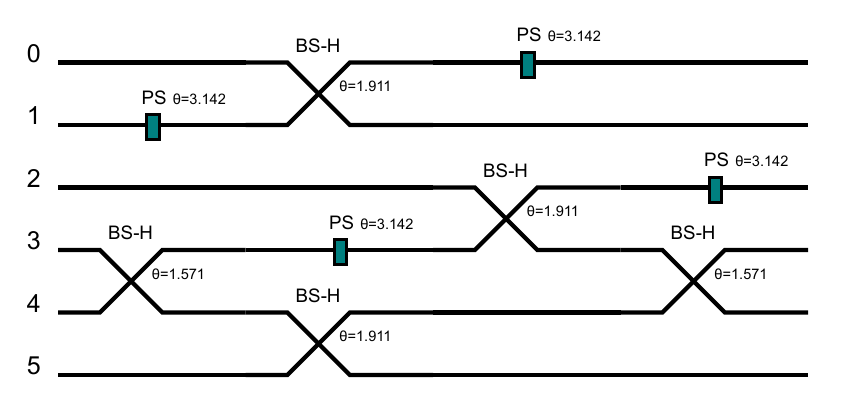}
    \caption{Visualization of the probabilistic photonic $\mathrm{CNOT}$ gate circuit constructed in Code~\ref{code:cnot}.}
    \label{fig:photonic_cnot}
\end{figure}

We utilize a 6-mode photonic quantum circuit to simulate this probabilistic $\mathrm{CNOT}$ gate, as implemented in Code~\ref{code:cnot}.
As illustrated in Fig.~\ref{fig:photonic_cnot}, among the six spatial modes, modes 0 and 5 serve as vacuum ancillas.
Modes 1 and 2 encode the first logical qubit (control), while modes 3 and 4 encode the second logical qubit (target) via the dual-rail scheme.
Correspondingly, the computational basis states map directly to the multiphoton Fock states:
\begin{equation}
    \begin{aligned}
        \ket{00}_L &\equiv \ket{1010}_{1,2,3,4}, \\
        \ket{01}_L &\equiv \ket{1001}_{1,2,3,4}, \\
        \ket{10}_L &\equiv \ket{0110}_{1,2,3,4}, \\
        \ket{11}_L &\equiv \ket{0101}_{1,2,3,4}.
    \end{aligned}
\end{equation}
A post-selection process is subsequently applied at the measurement stage: only events satisfying the coincidence conditions $n_1 + n_2 = 1$ and $n_3 + n_4 = 1$ are retained, ensuring that the retained measurement outcomes lie within the logical subspace.
As demonstrated by the probability amplitudes of 1/3 obtained in Code~\ref{code:cnot}, filtering out the invalid states through this post-selection yields a theoretical success probability of 1/9 for the implemented $\mathrm{CNOT}$ operation.

\subsubsection{Unitary Transformation with Clements Architecture}
Linear optical quantum computing (LOQC) relies on the precise implementation of arbitrary unitary transformations across multiple photonic modes~\cite{russell2017direct, reck1994experimental, carolan2015universal, bogaerts2020programmable, clements2016optimal}.
Among various approaches, the Clements architecture~\cite{clements2016optimal} stands out as an efficient and compact design for decomposing an arbitrary $N$-mode unitary matrix into a structured sequence of optical components, specifically beam splitters and phase shifters.

\begin{lstlisting}[caption={Implementation of the Clements architecture for unitary decomposition.}, label={code:clements}]
import deepquantum as dq
import torch

u = (
    torch.tensor(
        [
            [1, 0, 1, -1, 0, 0],
            [0, 1, 0, 0, 0, 2**0.5],
            [1, 0, 0, 1, 1, 0],
            [-1, 0, 1, 0, 1, 0],
            [0, 0, 1, 1, -1, 0],
            [0, 2**0.5, 0, 0, 0, -1],
        ]
    )
    / 3**0.5
)
# Decompose the unitary matrix
ud = dq.UnitaryDecomposer(u)
angle_dict = ud.decomp()[2]
clements = dq.Clements(nmode=6, init_state=[1, 0, 1, 0, 0, 0], cutoff=3)
# Encode the parameters into the 6-mode Clements architecture
data = clements.dict2data(angle_dict)
clements(data=data)
clements.draw()
\end{lstlisting}

\begin{figure}[!htbp]
    \centering
    \includegraphics[width=\linewidth]{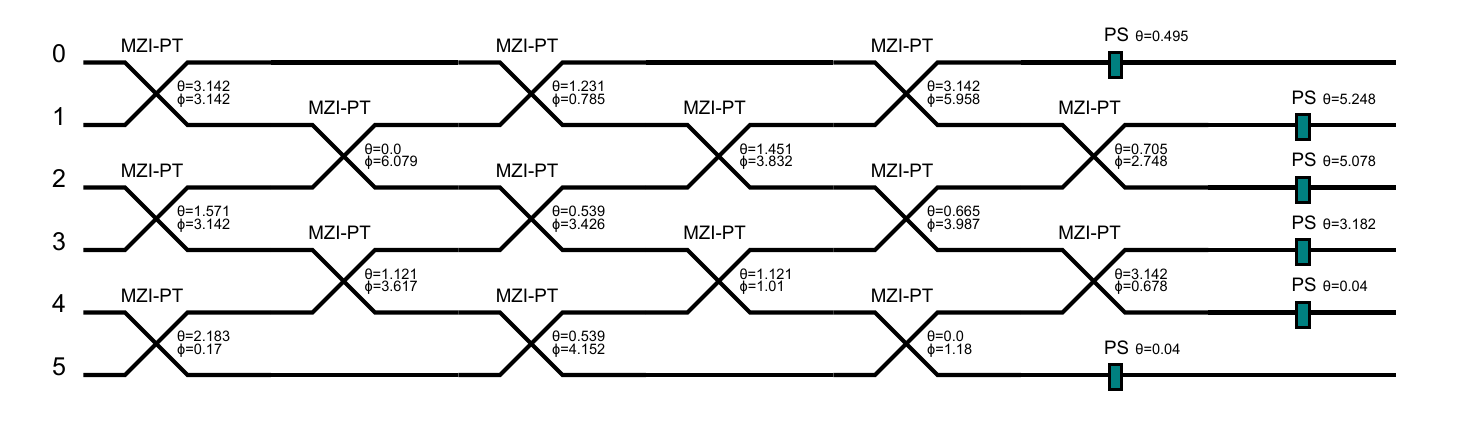}
    \caption{Visualization of the 6-mode Clements architecture circuit generated in Code~\ref{code:clements}.}
    \label{fig:clements}
\end{figure}

Code~\ref{code:clements} demonstrates the implementation of a 6-mode Clements architecture within DeepQuantum.
The \texttt{UnitaryDecomposer} class decomposes a target unitary matrix into the specific structural parameters required for the Clements architecture.
These parameters are subsequently encoded into the circuit using the \texttt{dict2data} method.
The resulting circuit architecture, along with the parameters for the constituent beam splitters and phase shifters, is visualized in Fig.~\ref{fig:clements}.

\subsubsection{Gaussian Boson Sampling}
GBS is a prominent CV quantum computational model that leverages the quantum properties of light to address complex sampling problems~\cite{hamilton2017gaussian, brod2019photonic}.
In a typical GBS setup, single-mode squeezed vacuum states are injected into a linear optical interferometer and subsequently detected in the Fock basis.
The task of sampling from the output probability distribution of a GBS device is widely believed to be classically intractable, establishing GBS as a prime candidate for demonstrating quantum computational advantage~\cite{hamilton2017gaussian, bulmer2022boundary, zhong2020quantum, zhong2021phase, madsen2022quantum}.
Furthermore, GBS offers significant algorithmic potential for solving practical problems that can be natively encoded into its architecture, such as finding dense subgraphs~\cite{arrazola2018using}, graph similarity analysis~\cite{schuld2020measuring}, and molecular docking~\cite{banchi2020molecular}.

DeepQuantum provides an intuitive interface for the direct construction of GBS circuits via the \texttt{GaussianBosonSampling} class, where users can explicitly define the input squeezing parameters and the linear optical interferometer.
Additionally, the framework offers a higher-level abstraction, the \texttt{GraphGBS} class, specifically tailored for tackling graph-related problems.
This functionality streamlines the workflow by automatically encoding a graph's adjacency structure into the corresponding GBS circuit and subsequently performing the sampling.

\begin{lstlisting}[caption={Setup of a basic GBS circuit.}, label={code:gbs}]
import deepquantum.photonic as dqp
import torch

# Define squeezing parameters and unitary transformation
squeezing = [1.0] * 6
unitary = torch.eye(6, dtype=torch.cfloat)
# Initialize the GBS circuit
gbs = dqp.GaussianBosonSampling(nmode=6, squeezing=squeezing, unitary=unitary)
gbs()
# Perform measurements with different detectors
result = gbs.measure(shots=1024, detector='pnrd')
print('pnrd:', result)
result = gbs.measure(shots=1024, detector='threshold')
print('threshold:', result)
# Draw the circuit
gbs.draw()
\end{lstlisting}

\noindent Output:
\begin{verbatim}
Using chain-rule method to sample the final states!
pnrd:{|000020>: 64, |200002>: 21, |000000>: 233, ...}
Using chain-rule method to sample the final states!
threshold:{|000110>: 27, |000010>: 40, ...}
\end{verbatim}

\begin{figure}[!htbp]
    \centering
    \includegraphics[width=\linewidth]{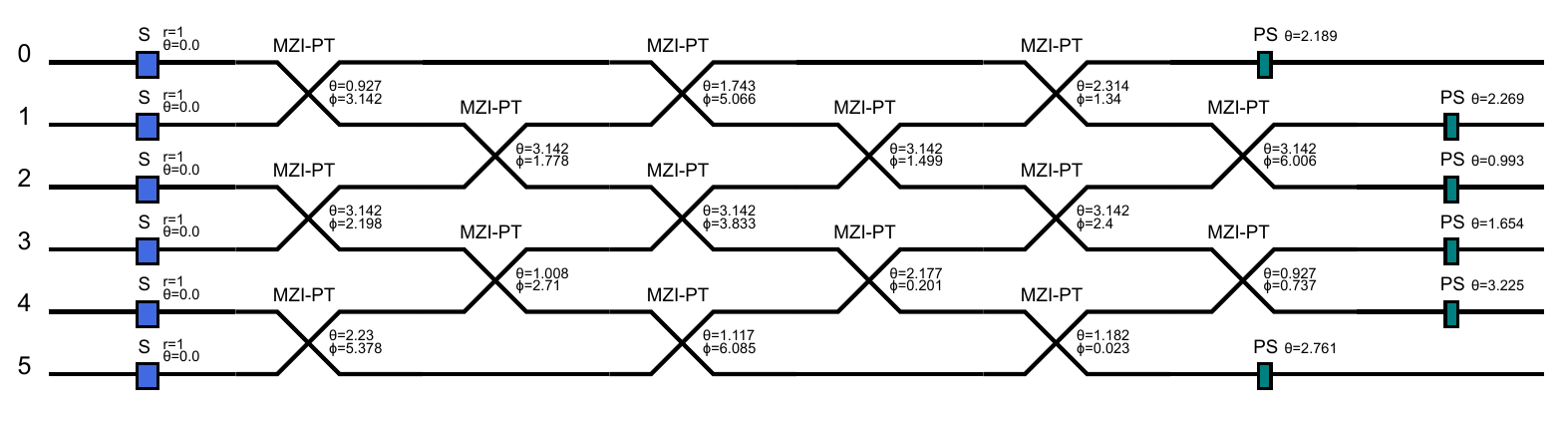}
    \caption{Visualization of the 6-mode GBS circuit constructed in Code~\ref{code:gbs}.}
    \label{fig:gbs}
\end{figure}

Code~\ref{code:gbs} illustrates the foundational construction of a GBS circuit.
In this configuration, six modes are initialized with identical single-mode squeezing operations and propagated through an identity interferometer.
Subsequently, the \texttt{measure()} method is invoked to simulate the detection process.
DeepQuantum supports both photon-number-resolving detectors and threshold detectors for Fock basis measurements, accessible via the \verb|detector='pnrd'| and \verb|detector='threshold'| arguments, respectively.
The resulting circuit architecture is shown in Fig.~\ref{fig:gbs}.

The \texttt{GraphGBS} class seamlessly embeds graph information into a GBS device.
Mathematically, this is achieved by applying the Takagi-Autonne decomposition to the graph's symmetric adjacency matrix $A$~\cite{bradler2018gaussian, oh2024quantum}:
\begin{equation}
    A = U \Lambda U^T,
\end{equation}
where $U$ is a unitary matrix and $\Lambda = \mathrm{Diag}(\lambda_1, \dots, \lambda_N)$ is a diagonal matrix containing non-negative singular values.
After an appropriate rescaling of the adjacency matrix to satisfy the physical squeezing constraints, the singular values determine the squeezing parameters, while $U$ specifies the interferometer.

\begin{lstlisting}[caption={Encoding a graph into a GBS device and performing measurement.},label={code:gbs_graph}]
# Define the adjacency matrix of the graph
adj_mat = torch.tensor(
    [
        [0.0, 1.0, 1.0, 0.0, 0.0, 0.0],
        [1.0, 0.0, 0.0, 1.0, 0.0, 1.0],
        [1.0, 0.0, 0.0, 0.0, 0.0, 0.0],
        [0.0, 1.0, 0.0, 0.0, 1.0, 1.0],
        [0.0, 0.0, 0.0, 1.0, 0.0, 0.0],
        [0.0, 1.0, 0.0, 1.0, 0.0, 0.0],
    ]
)
# Initialize the GBS circuit from the graph
gbs = dqp.GraphGBS(adj_mat=adj_mat, cutoff=2)
gbs()
sample = gbs.measure()
print(sample)
\end{lstlisting}

\noindent Output:
\begin{verbatim}
Using chain-rule method to sample the final
states!
{|000000>: 399, |000101>: 58, |010100>: 118, ...}
\end{verbatim}

Code~\ref{code:gbs_graph} demonstrates a GBS setup directly configured from a six-node graph.
Once the graph is encoded, the \texttt{GraphGBS} instance inherently functions as a standard \texttt{QumodeCircuit}, autonomously managing the internal quantum state evolution and subsequent sampling procedures.

\subsubsection{Preparation of EPR State with TDM}
Quantum entanglement is a fundamental resource for quantum information processing in both DV and CV systems~\cite{nielsen2010quantum, braunstein2005quantum}.
Among the most widely utilized entangled states are the Einstein-Podolsky-Rosen (EPR) states~\cite{einstein1935can} and Greenberger-Horne-Zeilinger (GHZ) states~\cite{greenberger1990bell}, both of which play crucial roles in quantum communication and quantum networking~\cite{reid2009colloquium}.

A significant breakthrough in CV entanglement generation was achieved using TDM architectures based on optical delay loops, pioneered by the Furusawa group~\cite{yokoyama2013ultra, yoshikawa2016invited, takeda2019demand}.
This scheme allows for the deterministic and flexible generation of large-scale entangled states, such as sequential EPR pairs and GHZ states.
Furthermore, it establishes the foundation for synthesizing large-scale one- and two-dimensional cluster states, marking a pivotal advance toward scalable universal measurement-based quantum computation~\cite{menicucci2006universal, gu2009quantum, menicucci2014fault}.

In the $x$-quadrature basis, the ideal CV EPR state, which corresponds to the limit of infinite squeezing, can be mathematically expressed as:
\begin{equation}
    \ket{\mathrm{EPR}}_{AB} \propto \int_{-\infty}^{\infty} \ket{x}_A \otimes \ket{x}_B \dd{x}.
\end{equation}
Consequently, a homodyne measurement of the $x$-quadrature on mode $A$ yielding an outcome $x_0$ instantaneously projects mode $B$ into an identical eigenstate, theoretically guaranteeing that an identical measurement on mode $B$ will yield the exact same outcome $x_0$.

\begin{figure}[!htbp]
    \centering
    \includegraphics[width=\linewidth]{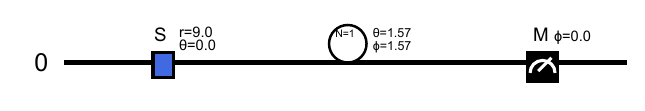}
    \caption{Two-mode TDM circuit for sequential EPR state preparation.}
    \label{fig:epr}
\end{figure}

The TDM circuit for generating sequential EPR entangled states is illustrated in Fig.~\ref{fig:epr}.
Initially, the beam splitter and phase shifter parameters of the delay loop are configured to $[\pi/2, \pi/2]$.
This specific setting effectively routes the initial vacuum state out of the loop directly to the detector, while simultaneously storing the first arriving squeezed vacuum state inside the delay loop.
Subsequently, the parameters are dynamically switched to $[\pi/4, 0]$, enabling a balanced interference between two consecutive squeezed vacuum states to create an EPR pair.
By periodically alternating this dynamic switching process, the circuit iteratively produces a continuous sequence of bipartite EPR states.

\begin{lstlisting}[caption={Simulation of TDM-based EPR state preparation.}, label={code:epr}]
import deepquantum as dq
import torch

cir = dq.QumodeCircuitTDM(nmode=1, init_state='vac')
cir.s(0, r=9.0)
cir.delay(0, ntau=1, inputs=[torch.pi/2, torch.pi/2], encode=True)
cir.homodyne_x(0)
data = torch.tensor([[torch.pi/2, torch.pi/2],
                     [torch.pi/4, 0.0]]).unsqueeze(0)
cir(data=data, nstep=13)
print(cir.samples[1:])
\end{lstlisting}

\noindent Output:
\begin{verbatim}
tensor([ 4112.4751,  4112.4751,
        -5751.3506, -5751.3506,
        -6159.6787, -6159.6787,
         526.8242,   526.8240,
        -6301.3857, -6301.3857,
        -5215.0078, -5215.0078])
\end{verbatim}

The realization of this sequential generation using DeepQuantum is demonstrated in Code~\ref{code:epr}.
Although the squeezing parameter significantly exceeds current experimental limits, it is intentionally employed here in the numerical simulation to approximate the infinitely sharp correlations characteristic of ideal EPR states.
Crucially, the numerical output is consistent with the underlying physical theory: after discarding the initial vacuum state via the \texttt{[1:]} slice, the consecutive $x$-quadrature homodyne measurements perfectly group into adjacent pairs of highly correlated values.
This strict numerical correlation supports the successful simulation of strongly correlated EPR-like pairs within the simulated TDM architecture.

\subsubsection{Preparation of Two-Dimensional Cluster State with TDM}
\begin{figure}[!htbp]
    \centering
    \includegraphics[width=\linewidth]{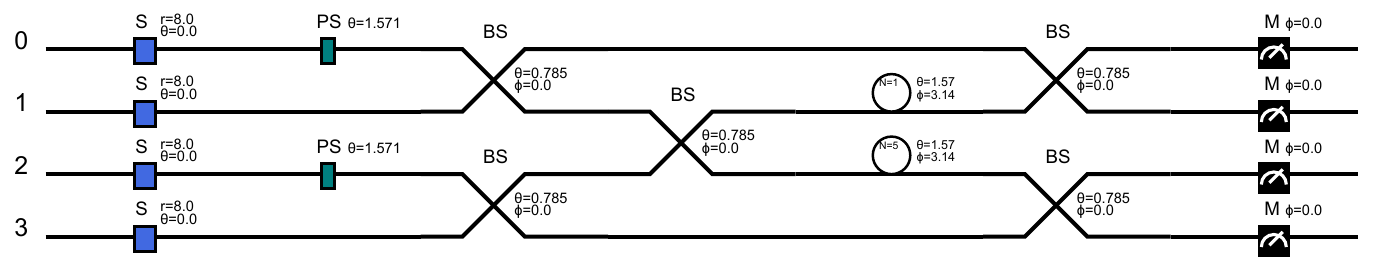}
    \caption{Schematic of the four-mode TDM circuit for 2D CV cluster state generation.}
    \label{fig:cluster}
\end{figure}

A landmark experimental realization of a large-scale, two-dimensional cluster state was reported in Ref.~\cite{asavanant2019generation}.
This was achieved through the generation of CV entanglement in a four-mode photonic quantum circuit featuring two asymmetric delay loops, as illustrated in Fig.~\ref{fig:cluster}.

The generation process begins with the preparation of four independent squeezed vacuum states.
At each time step, these states are directed into a network of three 50:50 beam splitters, where they interfere to entangle the modes.
Following this interaction, mode 1 is routed through a delay loop corresponding to a single time period ($\tau$), while mode 2 undergoes a significantly longer delay of $N$ periods ($N\tau$), with $N=5$ being a typical experimental configuration.
After the respective delays, the modes undergo a final layer of beam splitter interference.
This asymmetric delay architecture weaves the entanglement across both temporal and spatial domains, constructing a cylindrical 2D lattice.

The topology of the resulting cluster state is rigorously characterized by a set of nullifier relations.
For this specific 2D architecture, one such relevant nullifier operator $\hat{\delta}_k$ is defined as:
\begin{equation}
    \hat{\delta}_k \equiv \hat{x}_k^A + \hat{x}_k^B - \frac{1}{\sqrt{2}} \pqty{-\hat{x}^A_{k+1} + \hat{x}^B_{k+1} + \hat{x}^C_{k+N} + \hat{x}^D_{k+N}},
    \label{eq:nullifier}
\end{equation}
where $k$ indexes the temporal steps, and the superscripts $A, B, C, D$ correspond to the four distinct spatial modes (mapped to modes 0, 1, 2, and 3 in the simulation, respectively).
In the non-physical limit of infinite squeezing, the variance of this nullifier vanishes, i.e., $\expval*{(\Delta \hat{\delta}_k)^2} \to 0$.

\begin{figure}[!htbp]
    \centering
    \includegraphics[width=\linewidth]{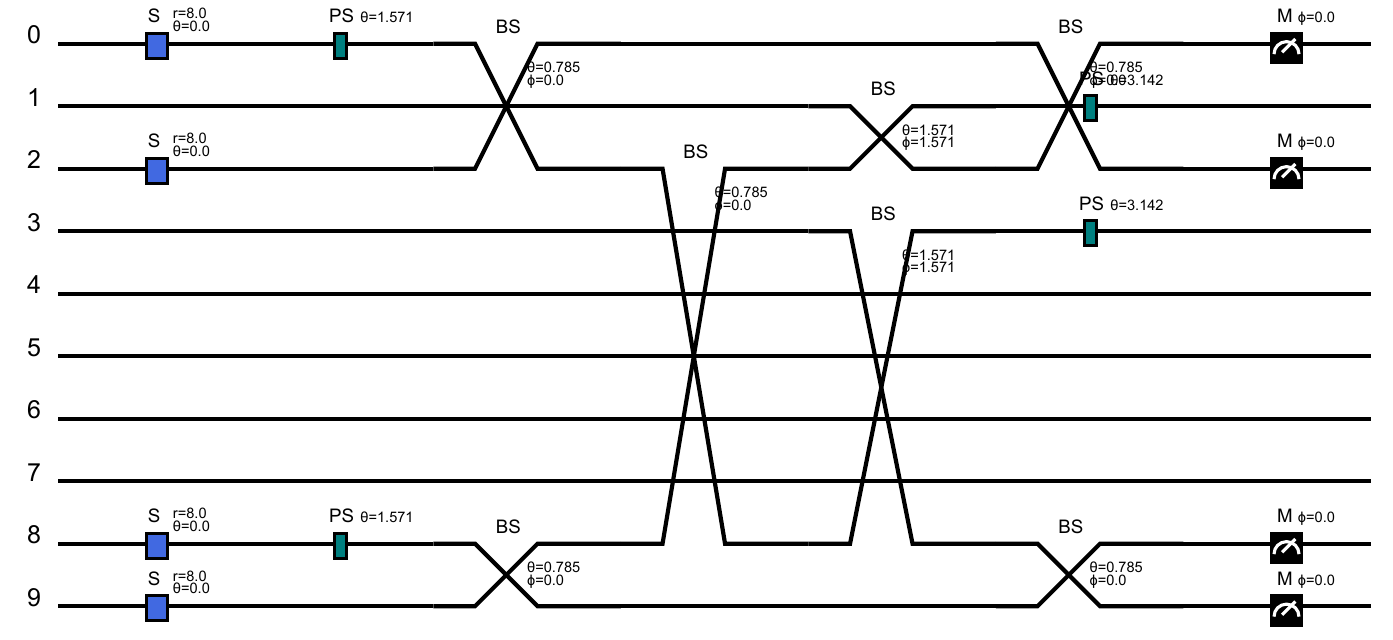}
    \caption{Equivalent unrolled spatial circuit corresponding to a single time step of the TDM 2D cluster state generation.}
    \label{fig:cluster_unroll}
\end{figure}

\begin{lstlisting}[caption={Simulation of the 2D cluster state preparation.}, label={code:cluster}]
import deepquantum as dq
import torch

cir = dq.QumodeCircuitTDM(4, 'vac', 3)
for i in range(4):
    cir.s(i, r=8.0)
cir.r(0, torch.pi/2)
cir.r(2, torch.pi/2)
cir.bs([0, 1], [torch.pi/4, 0])
cir.bs([2, 3], [torch.pi/4, 0])
cir.bs([1, 2], [torch.pi/4, 0])
cir.delay(1, ntau=1, inputs=[torch.pi/2, torch.pi])
cir.delay(2, ntau=5, inputs=[torch.pi/2, torch.pi])
cir.bs([0, 1], [torch.pi/4, 0])
cir.bs([2, 3], [torch.pi/4, 0])
for i in range(4):
    cir.homodyne_x(i, eps=1e-6)
cir.to(torch.double)
cir()
\end{lstlisting}

To intuitively understand the generation dynamics, the sequential TDM architecture can be conceptually unrolled into an expansive spatial circuit.
The equivalent spatial circuit, representing a single time step (\texttt{nstep=1}), is explicitly visualized in Fig.~\ref{fig:cluster_unroll}.
To translate this theoretical and experimental framework into a numerical simulation, Code~\ref{code:cluster} demonstrates the complete state preparation process using DeepQuantum.

\begin{lstlisting}[caption={Verification of the cluster state via nullifier evaluation.}, label={code:nullifier}]
cir(nstep=100)
samples = cir.samples.mT
nullifiers = []
for i in range(90):
    nullifier = samples[i][:2].sum() - 1 / 2**0.5 * (-samples[i + 1][0] + samples[i + 1][1] + samples[i + 5][2:].sum())
    nullifiers.append(nullifier)
nullifiers = torch.stack(nullifiers)
print('mean:', nullifiers.mean())
print('variance:', nullifiers.std() ** 2)
\end{lstlisting}

\noindent Output:
\begin{verbatim}
mean: tensor(-3.7434e-05, dtype=torch.float64)
variance: tensor(5.4542e-07, dtype=torch.float64)
\end{verbatim}

To confirm the correctness of the generated state, we evaluate the nullifier defined in Eq.~\eqref{eq:nullifier} utilizing the homodyne measurement samples across 100 consecutive time steps.
The verification procedure is detailed in Code~\ref{code:nullifier}.
Remarkably, the mean of the residuals is tightly bounded around zero, and the variance evaluates to $5.45 \times 10^{-7}$.
Because the theoretical variance of a nullifier scales proportionally with $e^{-2r}$, this numerical result matches the expected order of magnitude for the specified squeezing parameter $r=8.0$, thus supporting the successful simulation of the 2D cluster state.

\subsubsection{Generation of Non-Gaussian GKP State}
Non-Gaussian states are indispensable resources in CV quantum information processing.
They play a crucial role in enabling universal quantum computation, providing the necessary nonlinear operations to achieve quantum advantage, and facilitating robust quantum error correction.
Unlike Gaussian states, whose Wigner functions are nonnegative, many important non-Gaussian states exhibit localized negative regions in their phase-space Wigner representations, serving as rigorous signatures of quantum non-classicality.
Among these, the GKP state stands out as a foundational bosonic quantum error-correcting code designed to mitigate small displacement errors in phase space.

\begin{figure}[!htbp]
    \centering
    \includegraphics[width=\linewidth]{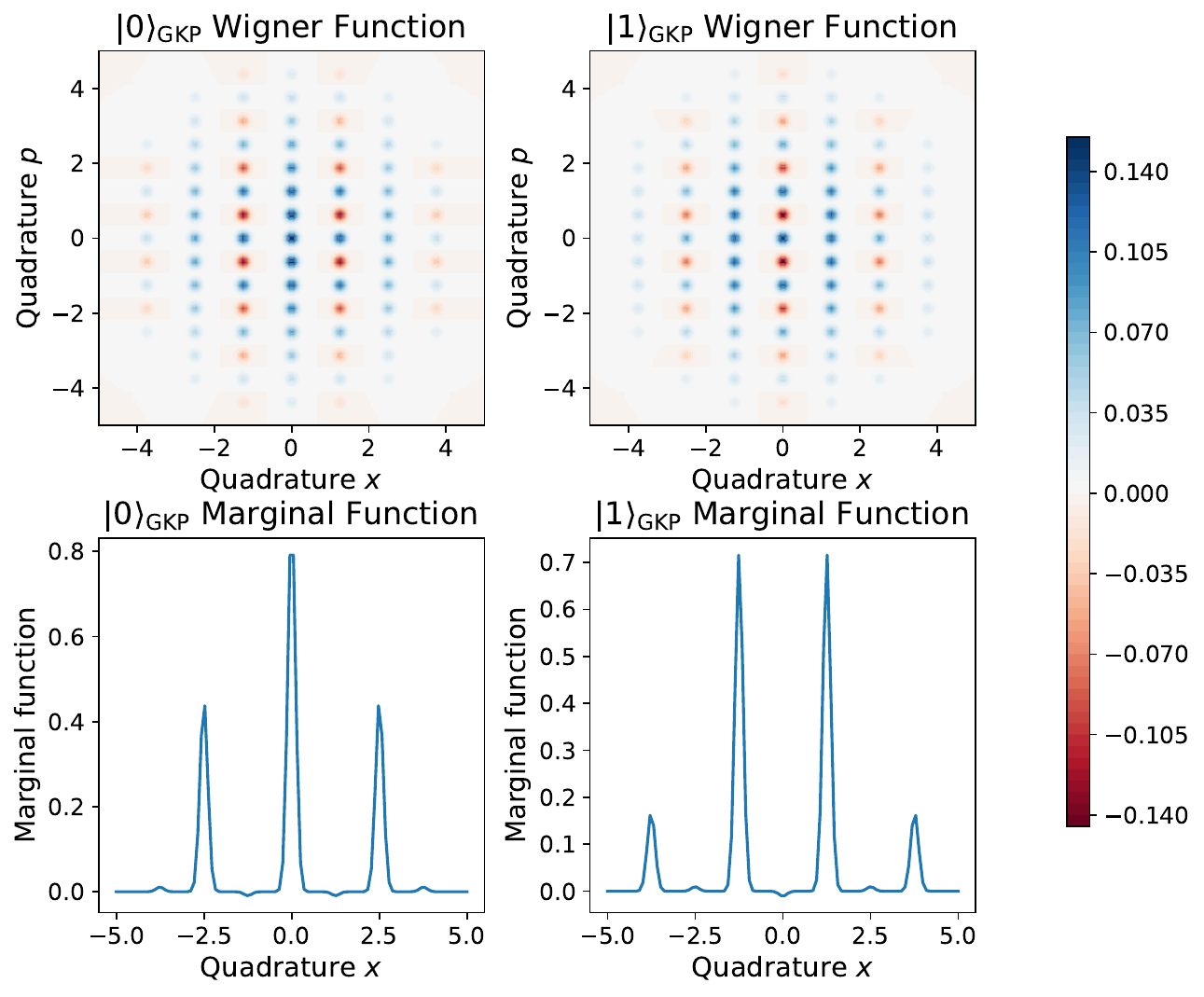}
    \caption{Wigner functions of finite-energy GKP logical states $\ket{0}_{\mathrm{GKP}}$ and $\ket{1}_{\mathrm{GKP}}$.}
    \label{fig:gkp_wigner}
\end{figure}

Because ideal GKP states correspond to an infinite grid of Dirac $\delta$-functions requiring infinite energy, physical implementations necessitate an energy constraint.
Mathematically, a finite-energy GKP state~\cite{matsuura2020equivalence} can be modeled by applying a Fock damping operator $e^{-\epsilon \hat{n}}$ to the ideal state~\cite{menicucci2014fault, bourassa2021fast}.
This non-unitary operation transforms the infinite $\delta$-grid into a superposition of finite-width Gaussian wave packets constrained by a broader global Gaussian envelope, as illustrated in Fig.~\ref{fig:gkp_wigner}.

Experimentally, approximate GKP states can be synthesized using two prominent optical approaches: the iterative breeding protocol~\cite{vasconcelos2010all, winnel2024deterministic} and the GBS-based post-selection method~\cite{larsen2025integrated}.

\begin{figure}[!htbp]
    \centering
    \includegraphics[width=\linewidth]{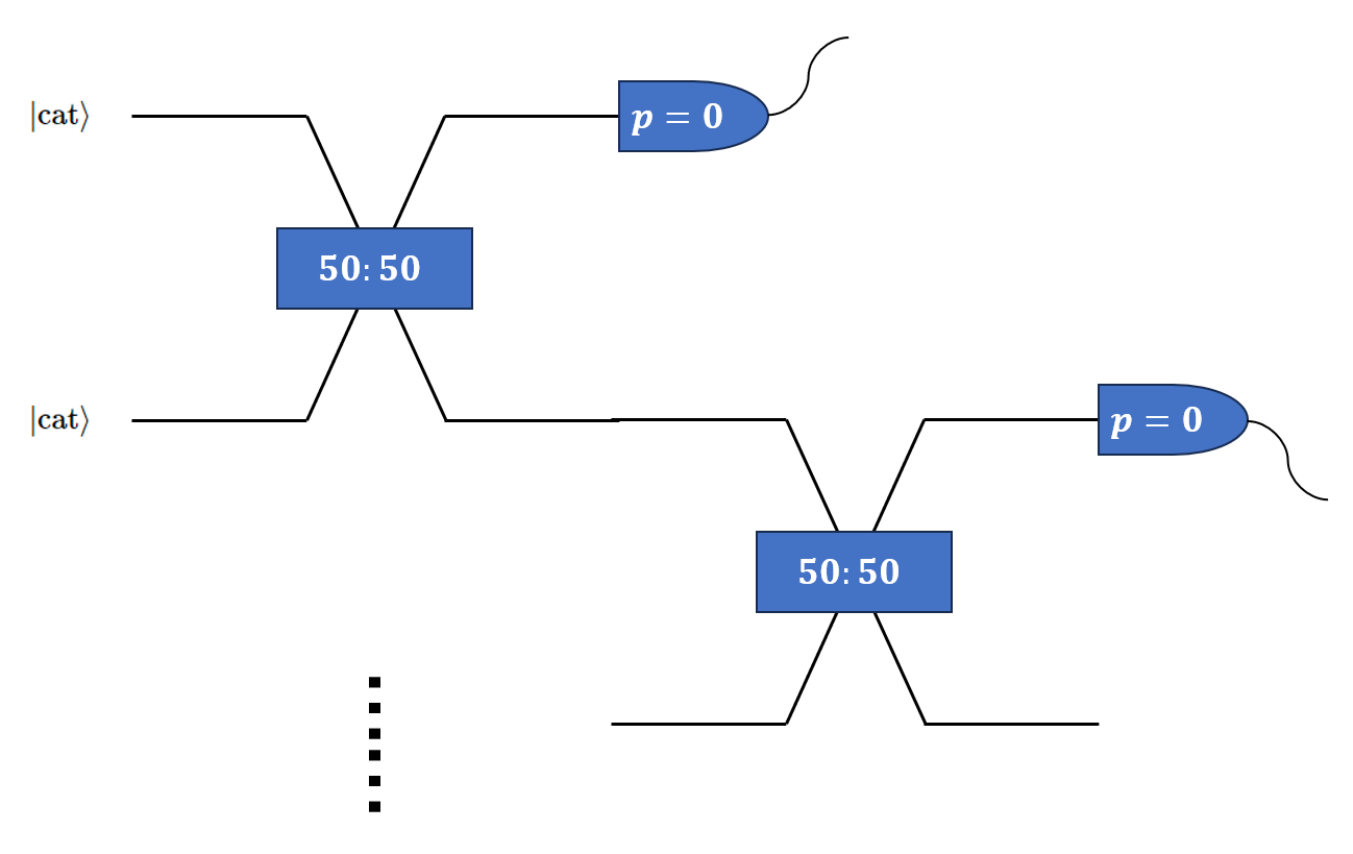}
    \caption{Schematic of the iterative breeding protocol for generating GKP states.
    The output of one breeding round serves as the resource state for the subsequent iteration.}
    \label{fig:breeding}
\end{figure}

In the breeding approach, two squeezed cat states are injected into a 50:50 beam splitter to entangle them.
Subsequently, a homodyne measurement is performed on the $p$-quadrature of the first mode.
By post-selecting measurement outcomes that are sufficiently close to zero, the unmeasured second mode collapses into a higher-quality approximation of a GKP state.
This process constitutes the first breeding stage.
To further refine the state, subsequent iterations employ the output states from previous rounds as new inputs, as schematically shown in Fig.~\ref{fig:breeding}.
As the number of breeding rounds increases, the resulting GKP state exhibits an increasingly pronounced lattice structure.

\begin{lstlisting}[caption={Simulation of the first breeding stage.}, label={code:breeding1}]
import deepquantum as dq
import numpy as np
import torch

r = 1.5
alpha = torch.tensor(2, dtype=torch.double)
alpha_prime = (np.cosh(r) + np.sinh(r)) * alpha / 2
cir = dq.QumodeCircuit(2, 'vac', backend='bosonic')
# Initialize plus cat state
cir.cat(wires=0, r=alpha_prime, theta=0, p=0)
cir.cat(wires=1, r=alpha_prime, theta=0, p=0)
cir.s(0, r)
cir.s(1, r)
cir.bs(wires=[0, 1], inputs=[np.pi/4, 0])
cir.homodyne_p(wires=0)
cir.to(torch.double)
cir()
sample = cir.measure_homodyne(shots=5000)
\end{lstlisting}

\begin{lstlisting}[caption={Conditioned post-selection of near-zero measurement results.}, label={code:breeding_select}]
for i in range(len(sample)):
    if abs(sample[i]) < 0.001:
        k = i
        break
idx = torch.tensor([1, 3])
bs_lst = [
    cir.state_measured[0][k][..., idx[:, None], idx],
    cir.state_measured[1][k][..., idx, :],
    cir.state_measured[2][k]
    ]
state_collapse = dq.BosonicState(bs_lst, nmode=1)
wigner = state_collapse.wigner(wire=0)
\end{lstlisting}

\begin{lstlisting}[caption={Simulation of the second breeding stage using previous outputs.}, label={code:breeding2}]
cir2 = dq.QumodeCircuit(2, [state_collapse, state_collapse], backend='bosonic')
cir2.bs(wires=[0, 1], inputs=[np.pi/4, 0])
cir2.homodyne_p(wires=0)
cir2.to(torch.double)
cir2()
sample2 = cir2.measure_homodyne(shots=1000)
for i in range(len(sample2)):
    if abs(sample2[i]) < 0.002:
        k = i
        break
idx = torch.tensor([1, 3])
bs_lst2 = [
    cir2.state_measured[0][k][...,idx[:, None], idx],
    cir2.state_measured[1][k][...,idx, :],
    cir2.state_measured[2][k]
    ]
state_collapse2 = dq.BosonicState(bs_lst2, nmode=1)
wigner = state_collapse2.wigner(wire=0)
\end{lstlisting}

We demonstrate a two-stage breeding process using DeepQuantum's Bosonic backend.
Code~\ref{code:breeding1} initializes the first stage with squeezed cat states and performs homodyne sampling.
Code~\ref{code:breeding_select} executes the crucial post-selection step, retaining only near-zero $p$-quadrature events and extracting the collapsed Bosonic state.
Finally, Code~\ref{code:breeding2} cascades the protocol by feeding the output of the first stage into the second breeding round to yield an even sharper GKP approximation.

\begin{figure}[!htbp]
    \centering
    \includegraphics[width=\linewidth]{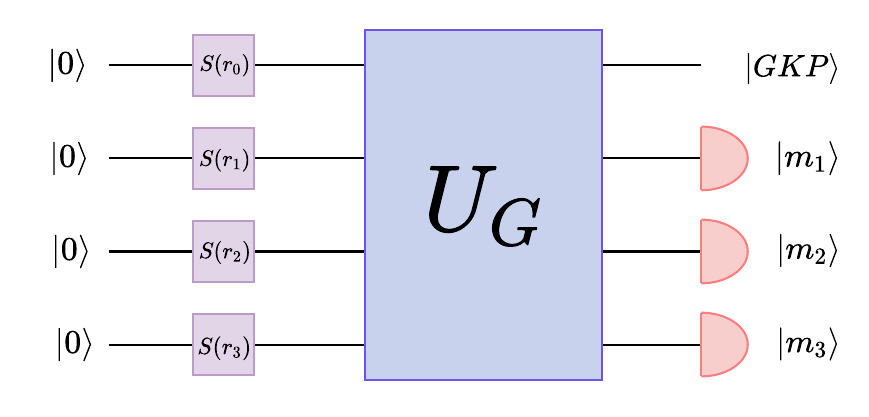}
    \caption{Schematic of GKP state preparation via the GBS-based post-selection method.}
    \label{fig:gkp_gbs}
\end{figure}

Alternatively, the GBS-based post-selection method offers another powerful pathway for generating GKP states, as depicted in Fig.~\ref{fig:gkp_gbs}.
In this architecture, four single-mode squeezed vacuum states are injected into a linear optical interferometer $U_G$, composed of beam splitters and phase shifters.
Photon-number-resolving detection is then implemented on modes 1, 2, and 3, yielding the photon counting outcome $(m_1, m_2, m_3)$.
By carefully designing the unitary transformation $U_G$ and heralding on a specific photon number pattern $(m_1, m_2, m_3)$, the unmeasured mode 0 collapses into the desired GKP state.

\begin{lstlisting}[caption={Preparation of the approximate GKP $\ket{1}$ state on a rectangular lattice via GBS post-selection.}, label={code:gkp_gbs}]
nmode = 4
cutoff = 50
squeeze = [1.0] * nmode
theta = [0.58, np.arctan(2**0.5), np.pi/4]
cir = dq.QumodeCircuit(nmode, 'vac', cutoff, backend='fock', basis=False)
for i in range(nmode):
    cir.s(wires=i, r=squeeze[i])
cir.ps(wires=1, inputs=np.pi/2)
cir.ps(wires=2, inputs=np.pi/2)
cir.ps(wires=3, inputs=np.pi/2)
cir.bs(wires=[0, 1], inputs=[theta[0], 0])
cir.bs(wires=[1, 2], inputs=[theta[1], 0])
cir.bs(wires=[2, 3], inputs=[theta[2], 0])
cir.to(torch.double)
state = cir()
# post-selection measurement
select = [3, 3, 3]
state0 = state[..., select[0], select[1], select[2]]
state0 = state0 / (state0.abs()**2).sum()**0.5
# Fock state to Wigner function
npoints = 100
fock = dq.FockState(state0, nmode=1, basis=False)
wigner = fock.wigner(wire=0, xrange=8, prange=8, npoints=npoints)
\end{lstlisting}

\begin{figure}[!htbp]
    \centering
    \includegraphics[width=\linewidth]{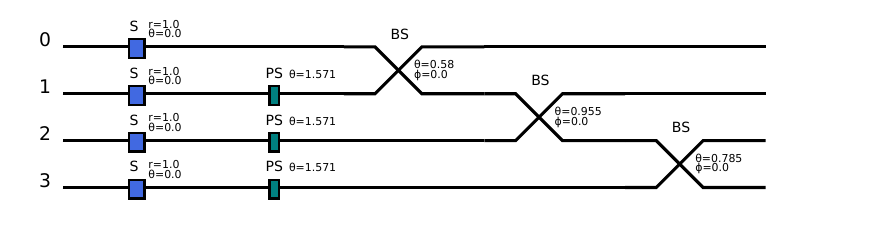}
    \caption{Configured GBS circuit for preparing the approximate GKP $\ket{1}$ state on a rectangular lattice.}
    \label{fig:gkp_gbs_circuit}
\end{figure}

\begin{figure}[!htbp]
    \centering
    \includegraphics[width=\linewidth]{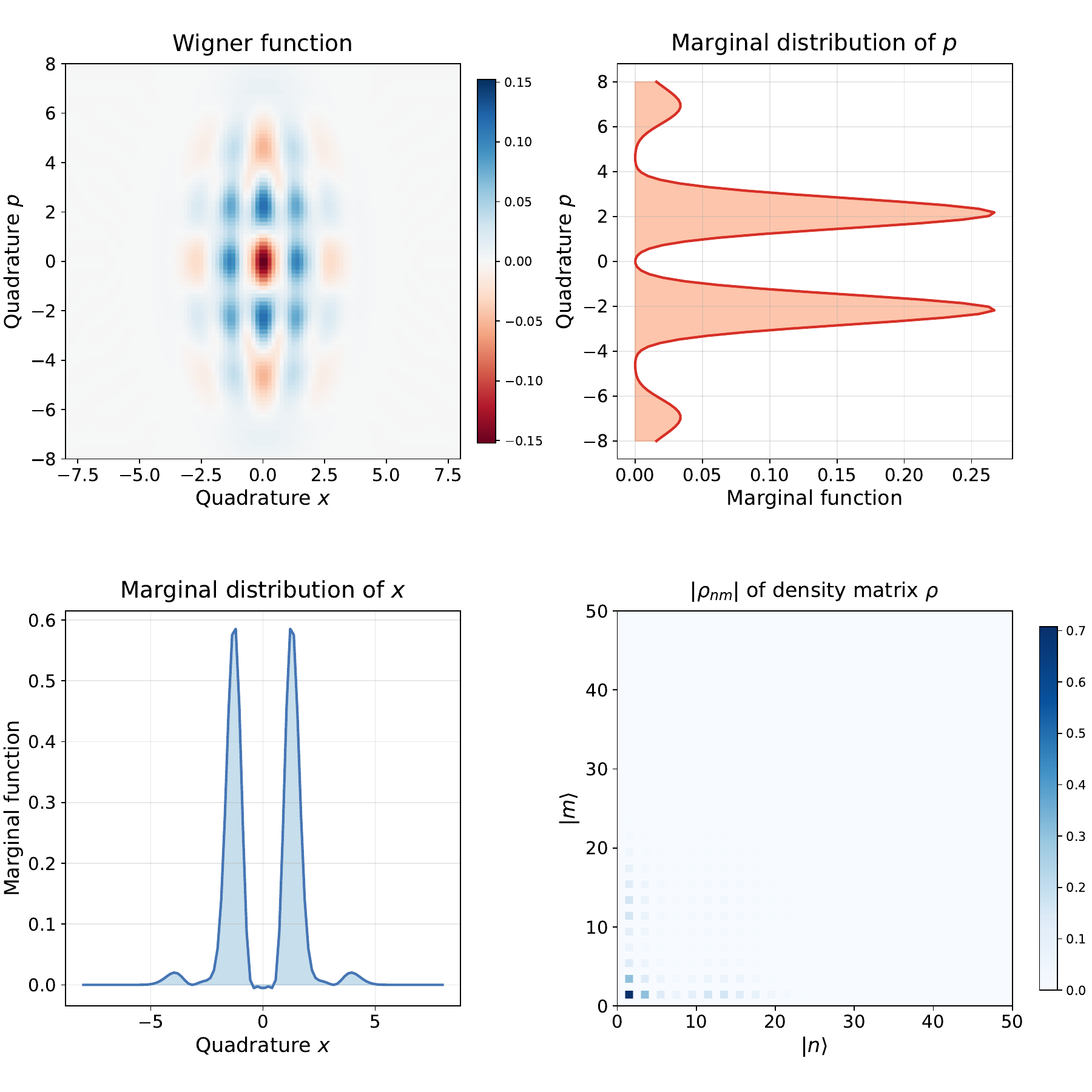}
    \caption{Visualizations of the approximate GKP $\ket{1}$ state on a rectangular lattice prepared via GBS post-selection.
    The panels display the Wigner function (top left), the marginal distributions (bottom left and top right), and the norm of the density matrix elements $\abs{\rho_{nm}}$ (bottom right).}
    \label{fig:gkp_gbs_results}
\end{figure}

Code~\ref{code:gkp_gbs} demonstrates the precise setup required to prepare the approximate GKP $\ket{1}$ state on a rectangular lattice utilizing the GBS framework with the Fock backend.
The detailed optical circuit configuration is visualized in Fig.~\ref{fig:gkp_gbs_circuit}.
Finally, the generation of the state is verified in Fig.~\ref{fig:gkp_gbs_results}, which displays the distinct Wigner negativity, the corresponding marginal distributions, and the density matrix elements of the heralded non-Gaussian state.

\section{Measurement-Based Quantum Computing}
\label{sec:mbqc}
\subsection{Background}
Measurement-based quantum computing uses cluster or graph states as resources, and operations are conducted as a sequence of measurements on certain nodes.
Each measurement outcome determines the basis of subsequent measurements and Pauli corrections, enabling information flow~\cite{raussendorf2001one, Raussendorf2003}.
MBQC can be conveniently implemented on photonic devices~\cite{Walther2005, Browne2005}, and has been applied in universal quantum algorithms~\cite{VandenNest2006}, fault-tolerant quantum computing~\cite{Raussendorf2006, Varnava2006, menicucci2014fault}, and blind quantum computing~\cite{Broadbent2009, Barz2012}.

\subsection{API Overview}
DeepQuantum provides a comprehensive framework for simulating MBQC, centered around the \texttt{Pattern} class.
It supports the end-to-end workflow encompassing pattern construction, transpilation, standardization, optimization, execution, and visualization, as elaborated below~\cite{danos2007measurement, sunami2022graphix}.

\begin{itemize}
    \item \textbf{Pattern construction.}
    An MBQC computation is characterized by a sequence of measurement calculus commands applied to a graph state.
    The graph state is constructed using the \texttt{Pattern.n($i$)} method to initialize node $i$, and the \texttt{Pattern.e($i$, $j$)} method to apply a $\mathrm{CZ}$ entanglement gate between nodes $i$ and $j$.
    The \texttt{Pattern.m($i$, $\alpha$, $\lambda$, $s$, $t$)} method measures node $i$ in the $\lambda \in \{\mathrm{XY}, \mathrm{YZ}, \mathrm{ZX}\}$ plane at a base angle $\alpha$.
    This angle is adaptively modified during execution based on the parity of the classical measurement outcomes residing in the feedforward domains $s$ (which governs the Pauli-$X$ dependency) and $t$ (which governs the Pauli-$Z$ dependency).
    Finally, \texttt{Pattern.x($i$, $s$)} and \texttt{Pattern.z($i$, $s$)} apply Pauli-$X$ and Pauli-$Z$ byproduct corrections on node $i$, conditioned on the classical outcomes in domain $s$.

    \item \textbf{Transpilation.}
    Alternatively, an MBQC pattern can be directly transpiled from a \texttt{QubitCircuit} via the \texttt{QubitCircuit.pattern()} method.
    The resulting \texttt{Pattern} is initially generated in a ``wild'' form, meaning that intermediate byproduct corrections have not yet been absorbed into the subsequent measurement angles.

    \item \textbf{Standardization and optimization.}
    Standardization is performed via the \texttt{Pattern.standardize()} method.
    This procedure systematically commutes byproduct corrections past entanglement and measurement commands, effectively absorbing them into the measurement angles.
    This eliminates the need for intermediate physical conditional gates, enabling a purely measurement-driven execution.
    Furthermore, optimization via the \texttt{Pattern.shift\_signals()} method minimizes the dependencies of measurement angles on the $t$-domain, thereby maximizing parallelizability and reducing the overall quantum depth of the pattern.

    \item \textbf{Execution.}
    Forward execution (invoked via \texttt{Pattern.forward()}) evaluates the MBQC pattern and returns the output as a \texttt{GraphState} object.
    Its \texttt{full\_state} attribute extracts the resulting normalized state vector, whose dimension natively corresponds to the Hilbert space of the remaining unmeasured nodes.
\end{itemize}

\begin{lstlisting}[caption={Workflow for constructing and optimizing an MBQC pattern from scratch.}, label={code:mbqc_construct}]
import deepquantum as dq
import numpy as np

# Initialization
pattern = dq.Pattern(nodes_state=[0, 1])
# Add nodes
pattern.n(2)
# Add edges
pattern.e(0, 2)
pattern.e(1, 2)
# Add measurements
pattern.m(node=0, angle=np.pi)
pattern.m(node=1, angle=np.pi, s_domain=[0])
pattern.n(3)
pattern.e(2, 3)
pattern.m(node=2, angle=np.pi, s_domain=[0], t_domain=[1])
# X correction
pattern.x(node=3, domain=[0, 1])
# Standardize and optimize the pattern
pattern.standardize()
pattern.shift_signals()
# Visualization
pattern.draw()
\end{lstlisting}

\begin{figure}[!htbp]
    \centering
    \includegraphics[width=0.48\linewidth]{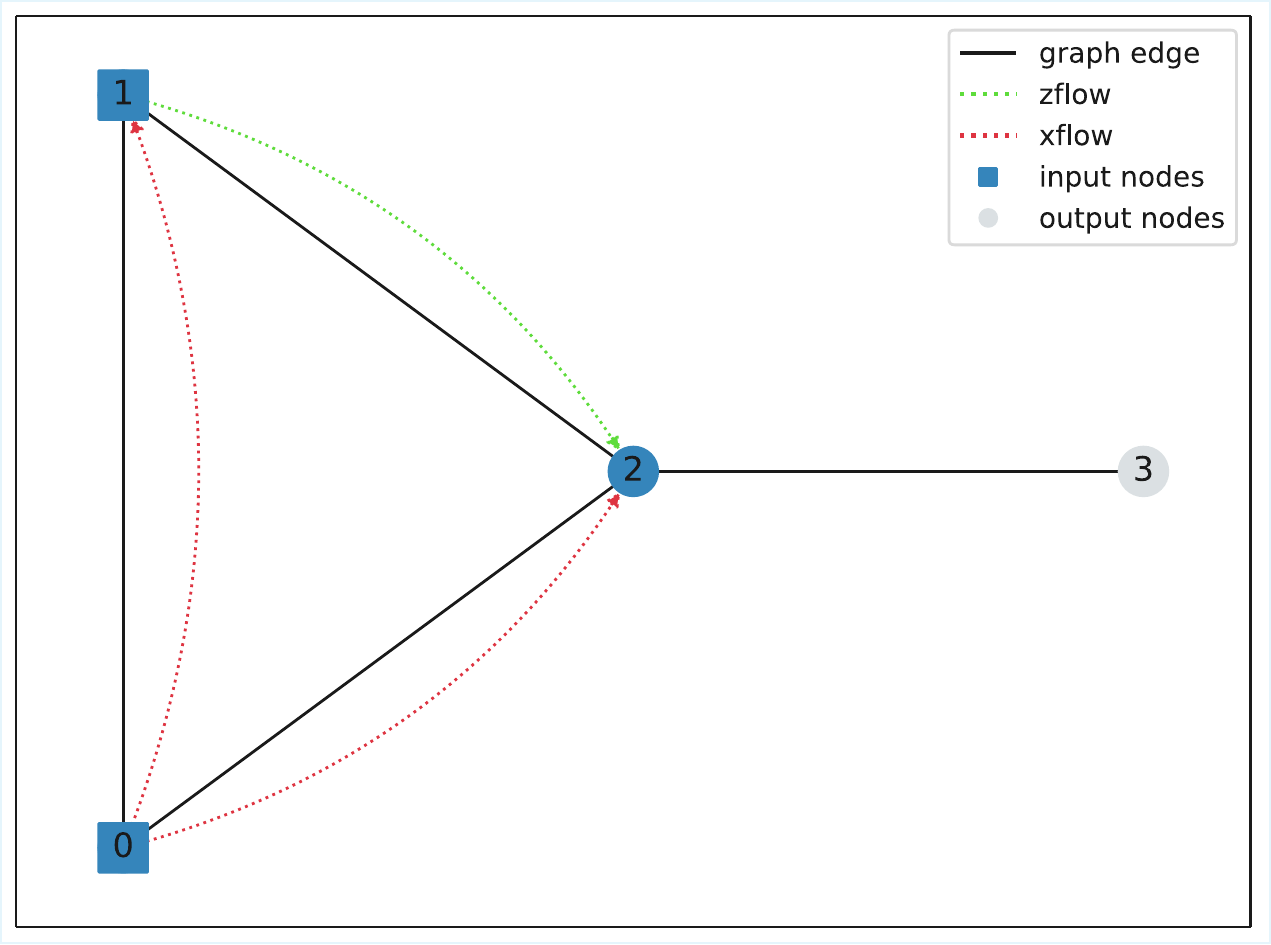}
    \includegraphics[width=0.48\linewidth]{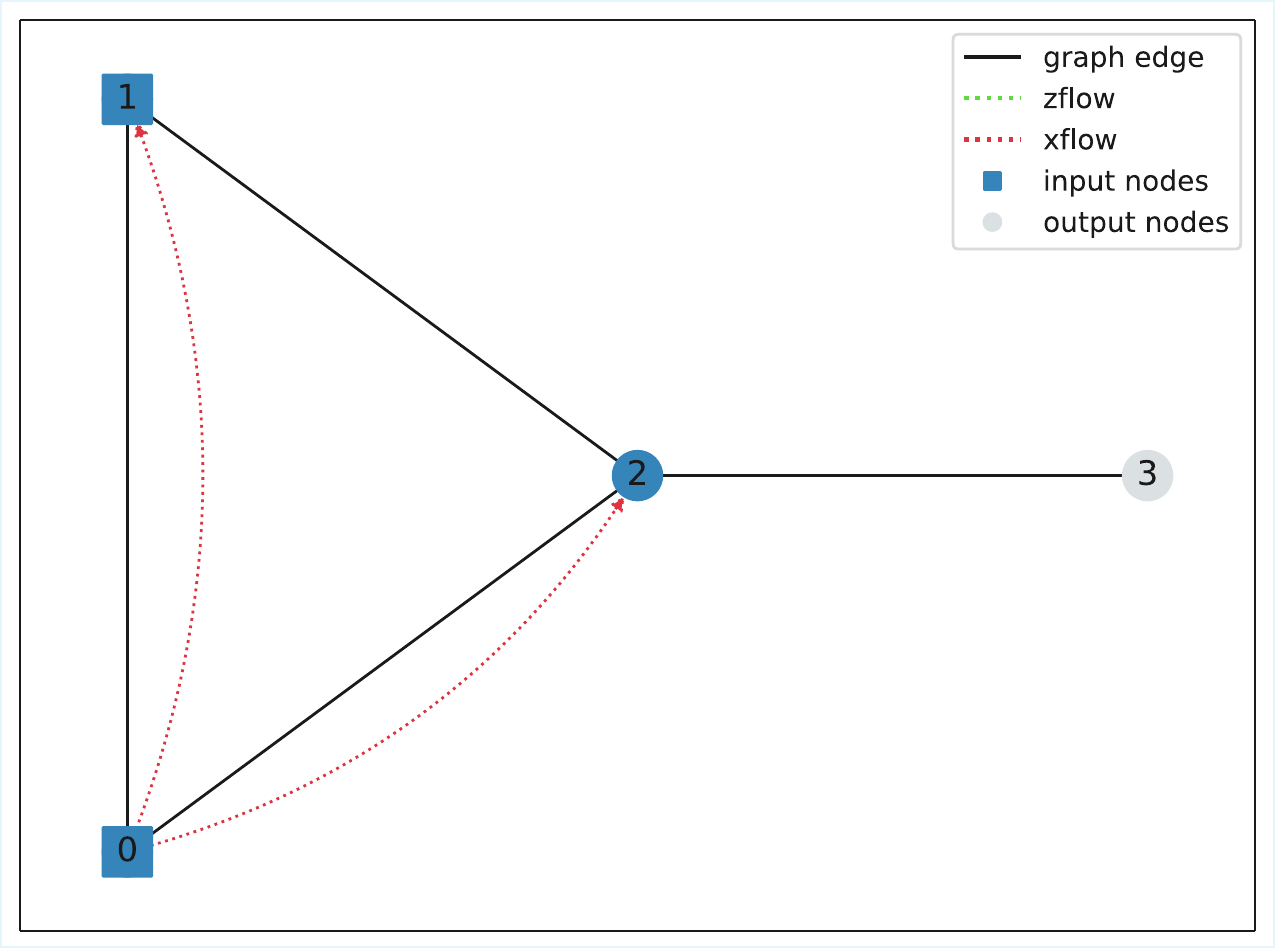}
    \caption{Visualization of the MBQC pattern constructed in Code~\ref{code:mbqc_construct}.
    Left: The standard pattern before signal shifting.
    Right: The optimized pattern after applying \texttt{shift\_signals()}, which simplifies feedforward dependencies and enhances parallelizability.}
    \label{fig:mbqc_opt_comparison}
\end{figure}

To demonstrate these capabilities, Code~\ref{code:mbqc_construct} presents a complete workflow for constructing, optimizing, and visualizing an MBQC pattern from scratch.
The visualizations of the pattern structure before and after the \texttt{shift\_signals()} optimization are compared in Fig.~\ref{fig:mbqc_opt_comparison}.

\begin{lstlisting}[caption={Workflow for transpiling a quantum circuit into an MBQC pattern.}, label={code:mbqc_transpile}]
# Construct a quantum circuit
circuit = dq.QubitCircuit(2)
circuit.h(0)
circuit.h(1)
circuit.cnot(0, 1)
circuit.draw()
# Transpile to a pattern
pattern = circuit.pattern()
# Standardize and optimize
pattern.standardize()
pattern.shift_signals()
pattern.draw()
\end{lstlisting}

\begin{figure}[!htbp]
    \centering
    \includegraphics[width=0.48\linewidth]{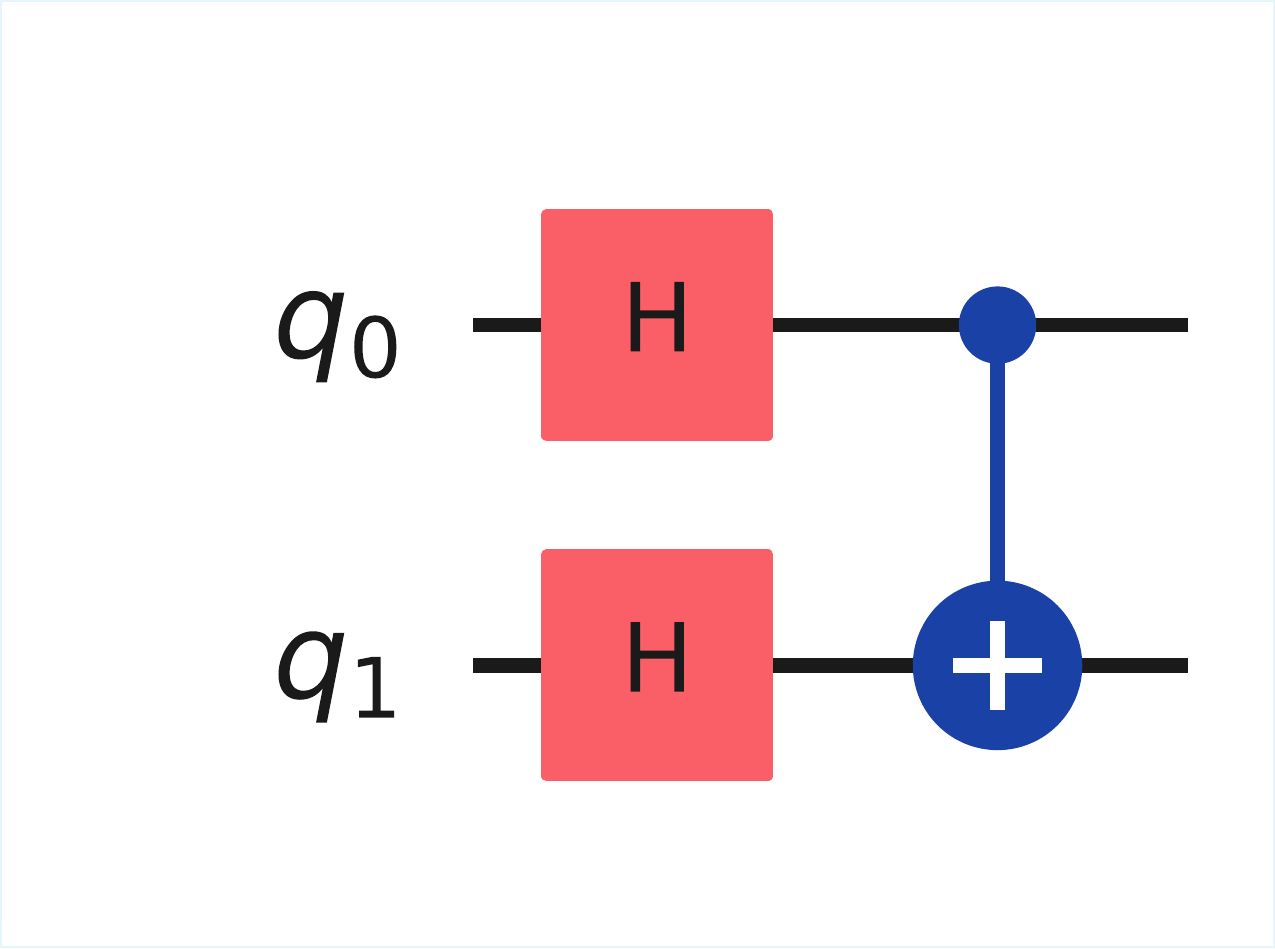}
    \includegraphics[width=0.48\linewidth]{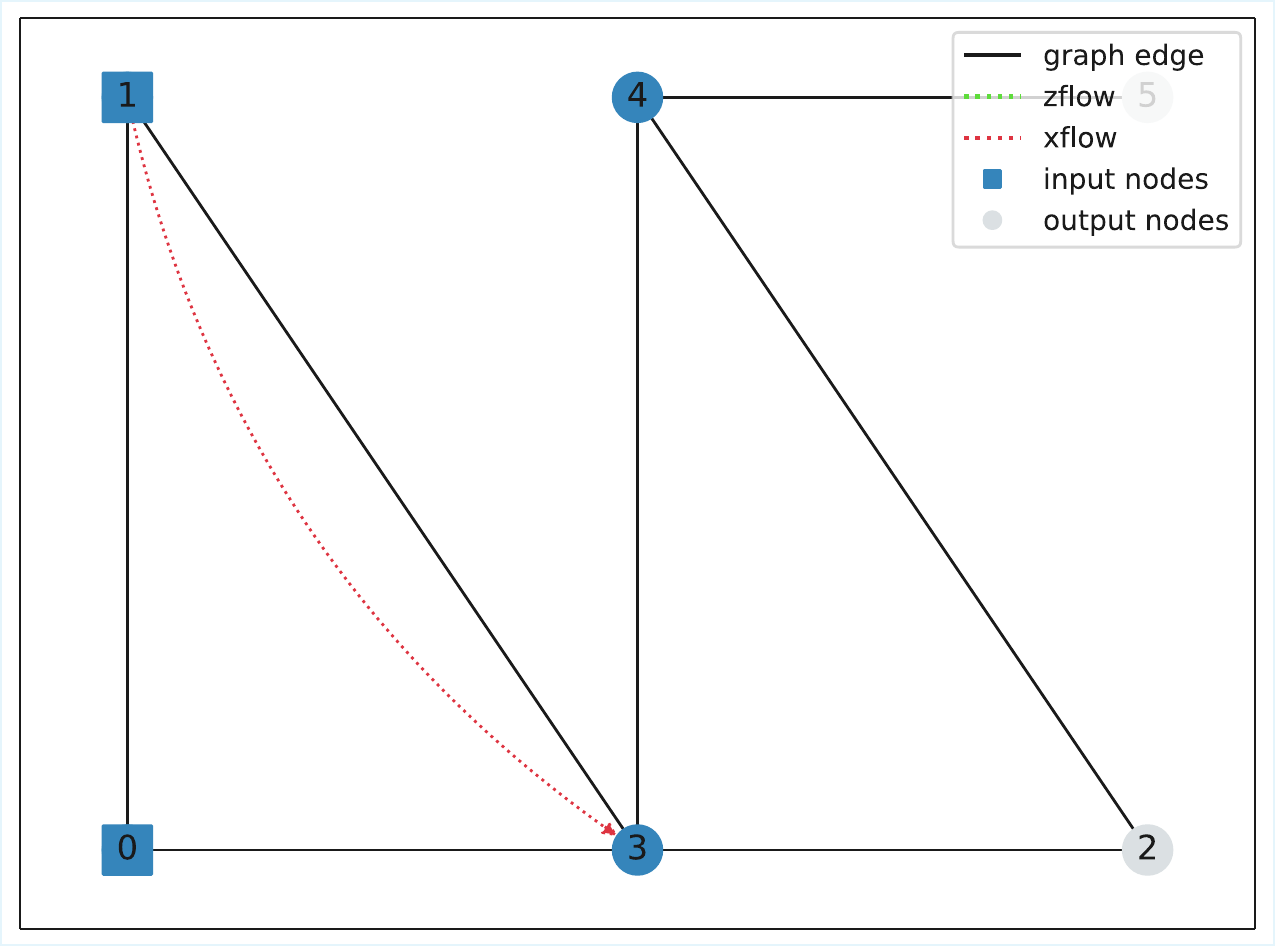}
    \caption{Comparison between a 2-qubit gate-based circuit (left) and its equivalent optimized MBQC measurement pattern (right), as generated in Code~\ref{code:mbqc_transpile}.}
    \label{fig:mbqc_transpile_comparison}
\end{figure}

For hybrid algorithm development, users often prefer designing models within the conventional gate-based circuit paradigm.
Code~\ref{code:mbqc_transpile} demonstrates how a \texttt{QubitCircuit} can be natively transpiled into a \texttt{Pattern}, followed by standardization and optimization.
The structural correspondence between the original \texttt{QubitCircuit} and its fully optimized MBQC \texttt{Pattern} is illustrated in Fig.~\ref{fig:mbqc_transpile_comparison}.

\subsection{Applications}
\begin{figure}[!htbp]
    \centering
    \includegraphics[width=\linewidth]{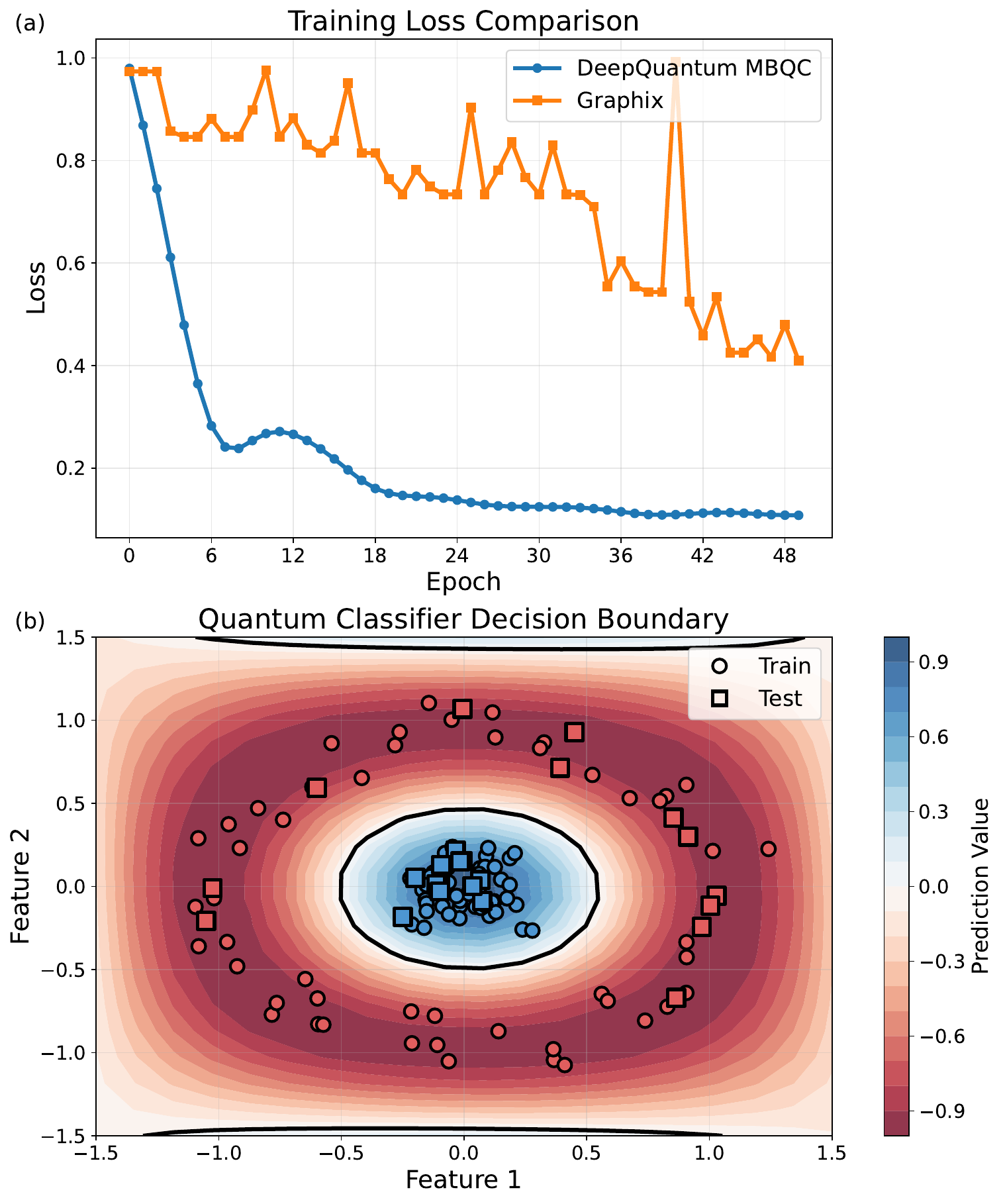}
    \caption{Performance of the quantum classifier with data re-uploading.
    (a) Loss convergence curves comparing DeepQuantum with Graphix.
    (b) Learned decision boundary showing the classification of training and test samples.}
    \label{fig:mbqc_training_results}
\end{figure}

In DeepQuantum, a variational quantum circuit can be transpiled into a variational MBQC pattern that preserves auto-differentiation.
In addition, data can be encoded into measurement angles using the \texttt{encode=True} flag.
Batch processing and data re-uploading are also supported.
Below, we train a quantum classifier\footnote{Following from the example presented at \url{https://graphix.readthedocs.io/en/v0.3.0/gallery/qnn.html\#sphx-glr-gallery-qnn-py}} whose quantum circuit ansatz employs the data-reuploading strategy.
Data re-uploading enables the use of fewer qubits and shallower circuits to perform complex tasks.
The classifier performs binary classification of the circles dataset, which is non-linearly separable.
Each input vector, consisting of two features, is zero-padded to three to accommodate three rotation gates.
In the $l$-th of 4 layers, the variational parameters $\boldsymbol{\theta}_l$, which are subsequently transpiled into the MBQC pattern, are given by
\begin{equation}
    \boldsymbol{\theta}_l = \boldsymbol{\beta}_l + \boldsymbol{\alpha}_l \odot \boldsymbol{x},
\end{equation}
where $\boldsymbol{\alpha}_l, \boldsymbol{\beta}_l$ are trainable parameters, $\odot$ denotes element-wise multiplication, and the input features $\boldsymbol{x}$ are re-uploaded across different layers.
We minimize the MSE loss
\begin{equation}
    {\cal L}=\frac{1}{N} \sum_{i=1}^N \abs{y^{(i)}-\expval{Z\otimes Z}_{\boldsymbol{\theta}}^{(i)}}^2.
\end{equation}
The training loss history and final decision boundary are shown in Fig.~\ref{fig:mbqc_training_results}, demonstrating the auto-differentiation advantages of DeepQuantum over Graphix, which lacks native auto-differentiation capabilities.

\section{Large-Scale Simulations}
\label{sec:large-scale}
\subsection{Tensor Network and Matrix Product State}
Tensor networks provide a powerful mathematical framework to overcome the exponential complexity inherent in large-scale quantum many-body systems and quantum circuits.
They have become indispensable tools for quantum circuit simulation~\cite{vidal2003efficient, Vidal2007, Orus2008, Shi2006}, optimization~\cite{Haegeman2016, Markov2008}, and entanglement analysis~\cite{Evenbly2009, Brandao2013, Eisert2010}.

A tensor can be represented as a multidimensional array of numbers defined with respect to a specific basis.
Mathematically, an order-$r$ tensor $T \in \mathbb{C}^{d_1 \times \dots \times d_r}$ is completely characterized by its scalar components $T_{i_1, \dots, i_r} \in \mathbb{C}$.
Here, each index $i_k$ can take values from 1 to $d_k$, where $d_k$ represents the local dimension of that particular index.
Following standard tensor network conventions, we will frequently use the indexed components to refer to the tensor as a whole.

 \begin{figure}[!htbp]
    \centering
    \includegraphics[width=\linewidth]{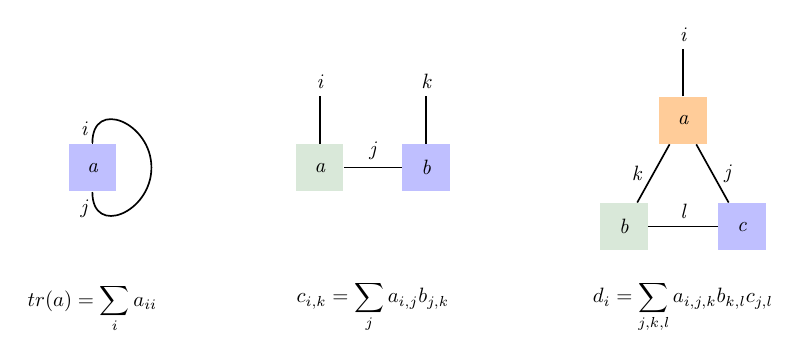}
    \caption{Penrose graphical notation for basic tensor contractions.}
    \label{fig:tensor_contraction}
\end{figure}

Scalars, vectors, and matrices correspond to order-0, order-1, and order-2 tensors, respectively.
In the Penrose graphical notation~\cite{Penrose1971}, an order-$r$ tensor is visually depicted as a geometric shape (e.g., a block or a circle) with $r$ emergent legs (lines), where each leg corresponds to one of the tensor's indices.
The fundamental algebraic operation within a tensor network is tensor contraction, which generalizes standard matrix multiplication.
Tensor contraction involves summing over one or more shared indices between two or more tensors.
In the graphical notation, this operation is elegantly expressed by connecting the legs corresponding to the shared indices to be summed over.
As illustrated in Fig.~\ref{fig:tensor_contraction}, this visual language simplifies the representation and manipulation of complex algebraic operations, such as traces and matrix multiplications.

The primary computational challenge in simulating large quantum systems is the so-called ``curse of dimensionality''.
The dimension of the Hilbert space for a system of $N$ qubits grows exponentially with $N$.
Specifically, an arbitrary pure state of $N$ qubits $\ket{\psi}$ is described by a state vector in a $2^N$-dimensional space:
\begin{equation}
    \ket{\psi} = \sum_{i_1, \dots, i_N} c_{i_1, \dots, i_N} \ket{i_1 \dots i_N},
\end{equation}
where each physical index $i_k \in \{0,1\}$.

 \begin{figure}[!htbp]
    \centering
    \includegraphics[width=\linewidth]{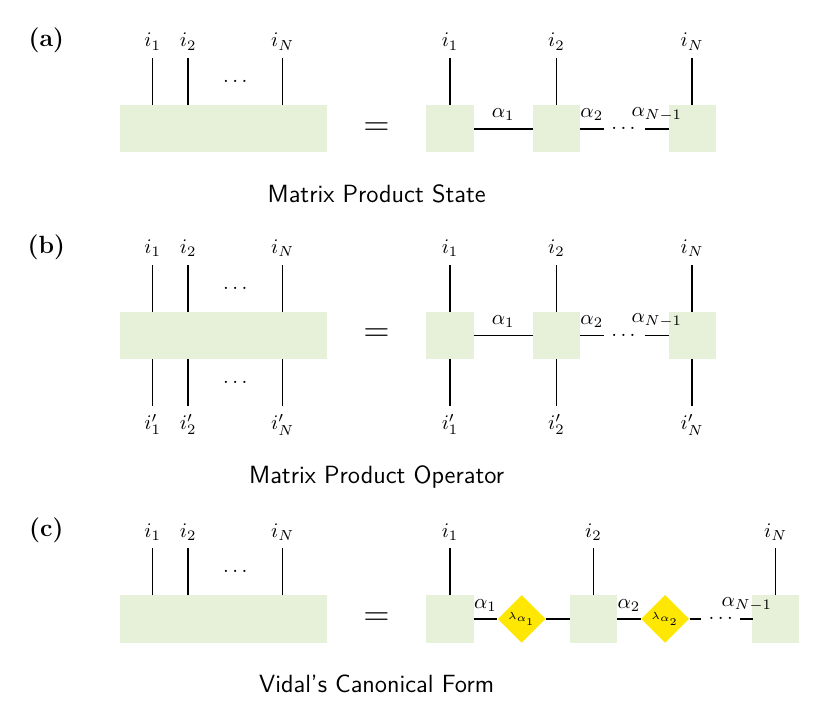}
    \caption{Graphical representations of 1D tensor networks.
    (a) An MPS with open boundary conditions.
    (b) An MPO, where each local tensor possesses both an input and an output physical leg.
    (c) An MPS expressed in Vidal's canonical form, explicitly showing the diagonal singular value matrices (diamonds) on the virtual bonds.}
    \label{fig:mps_mpo}
\end{figure}

In this context, the multidimensional amplitude $c_{i_1, \dots, i_N}$ is mathematically equivalent to an order-$N$ tensor comprising $2^N$ components.
Since directly storing and manipulating this tensor becomes computationally intractable for even a modest number of qubits, we employ the MPS formalism to provide an efficient approximation.
An MPS decomposes the global tensor into a contracted network of smaller, lower-order local tensors.
Assuming open boundary conditions (OBC), this decomposition reads~\cite{Cirac2021}:
\begin{equation}
    c_{i_1, \dots, i_N} = \sum_{\alpha_1=1}^{\chi}\cdots\sum_{\alpha_{N-1}=1}^{\chi} \Gamma_{1,\alpha_1}^{[1] i_1} \Gamma_{\alpha_1,\alpha_2}^{[2] i_2} \cdots \Gamma_{\alpha_{N-1},1}^{[N] i_N}.
\end{equation}
Here, each tensor $\Gamma^{[k]}$ represents an order-3 tensor, and the parameter $\chi$ is the bond dimension, which systematically controls the truncation error and approximation accuracy.
The virtual indices $\alpha_k$ are internal to the network and are contracted over, while each open physical index $i_k$ corresponds to the local physical state at site $k$.
For a qubit system, the physical dimension is $d=2$.
For a system in Fock space, $d$ is determined by the local photon-number cutoff.
The corresponding tensor diagram is depicted in Fig.~\ref{fig:mps_mpo}(a).

The MPS framework for pure states naturally extends to mixed states and operators via the matrix product operator (MPO) representation, which is essential for encoding Hamiltonians and other observables.
This is accomplished by augmenting each local tensor with a second physical index, transforming the order-3 MPS tensor $\Gamma_{\alpha_{k-1},\alpha_k}^{[k] i_k}$ into an order-4 MPO tensor $\Gamma_{\alpha_{k-1},\alpha_k}^{[k] i_k,i_k^\prime}$.
The indices $i_k$ and $i_k^\prime$ serve as the input and output physical legs, respectively, enabling the MPO to act as a linear map on the MPS space, as shown in Fig.~\ref{fig:mps_mpo}(b).

The MPS representation of a given quantum state is inherently not unique due to ``gauge freedom''.
For instance, inserting an invertible matrix and its inverse on any virtual bond leaves the global physical state invariant.
Vidal's canonical form~\cite{vidal2003efficient} eliminates this ambiguity by imposing specific orthogonality conditions, yielding a canonical representation up to residual freedoms associated with degeneracies or phase conventions (see Fig.~\ref{fig:mps_mpo}(c)):
\begin{equation}
    c_{i_1, \dots, i_N} = \sum_{\alpha_1=1}^{\chi}\cdots\sum_{\alpha_{N-1}=1}^{\chi} \Gamma_{1,\alpha_1}^{[1] i_1} \lambda^{[1]}_{\alpha_1} \Gamma_{\alpha_1,\alpha_2}^{[2] i_2} \lambda^{[2]}_{\alpha_2} \cdots \Gamma_{\alpha_{N-1},1}^{[N] i_N}.
\end{equation}

To construct an MPS from a dense state vector, we begin with the global coefficient tensor $c_{i_1, \dots, i_N}$.
The first step involves reshaping this tensor into a matrix $M_1$ of dimension $d \times d^{N-1}$.
This is achieved by treating the first physical index $i_1$ as the row index, and grouping the remaining $N-1$ indices into a composite column index.
Subsequently, a singular value decomposition (SVD) is applied to $M_1$:
\begin{equation}
    c_{i_1, i_2 \dots i_N} = \sum_{\alpha_1=1}^{\chi} \Gamma_{1,\alpha_1}^{[1] i_1} \lambda^{[1]}_{\alpha_1} (V^{\dagger})^{[1]}_{\alpha_1,i_2 \dots i_N}.
\end{equation}
This process is then iteratively executed.
For the second step, the right unitary matrix $(V^{\dagger})^{[1]}$ obtained previously is reshaped into a new matrix $M_2$ with dimension $d\chi \times d^{N-2}$, where the composite indices $(\alpha_1, i_2)$ constitute the row.
An SVD is subsequently applied to $M_2$, yielding the next local MPS tensor $\Gamma^{[2]}$ and the corresponding singular value matrix $\lambda^{[2]}$.
This sequence of reshaping and decomposing is repeated site by site until the entire state is fully factorized into the canonical MPS format.

The singular values $\lambda^{[k]}_{\alpha_k}$ obtained at each spatial step $k$ carry profound physical significance.
They correspond exactly to the Schmidt coefficients for the bipartition of the quantum system into subsystem $A$ (sites 1 to $k$) and subsystem $B$ (sites $k+1$ to $N$).
Consequently, they directly determine the bipartite entanglement entropy across this cut.

In practical large-scale simulations, we often enforce a global maximum bond dimension $\chi_{\mathrm{max}}$, dynamically truncating small singular values after each SVD operation.
To exactly represent a generic highly entangled $N$-site state with local dimension $d$, one may need a bond dimension as large as $\chi_{\mathrm{max}} \sim d^{\lfloor N/2 \rfloor}$.
By contrast, the ground states of many 1D local quantum many-body systems that satisfy the entanglement area law can often be efficiently approximated with much smaller bond dimensions~\cite{Verstraete2006}.
This physically motivated truncation drastically reduces the memory storage cost from $\mathcal{O}(d^N)$ to $\mathcal{O}(N d \chi^2)$ for an open-boundary MPS with a roughly uniform bond dimension $\chi$.

DeepQuantum natively supports simulations using the MPS representation for both the \texttt{QubitCircuit} and \texttt{QumodeCircuit} classes.
To enable this functionality, users simply set the \texttt{mps=True} flag during circuit initialization.
The maximum bond dimension, which fundamentally governs the simulation's precision and memory consumption, is explicitly configured using the \texttt{chi} argument.

\begin{lstlisting}[caption={Executing a 20-qubit circuit utilizing the MPS representation.}, label={code:mps_qubit}]
import deepquantum as dq

cir = dq.QubitCircuit(20, mps=True, chi=7)
cir.cnot_ring()
mps_state = cir()
for state in mps_state:
    print(state.shape)
\end{lstlisting}

\noindent Output:
\begin{verbatim}
torch.Size([1, 2, 2])
torch.Size([2, 2, 4])
torch.Size([4, 2, 7])
torch.Size([7, 2, 7])
......
torch.Size([2, 2, 1])
\end{verbatim}

For instance, the workflow presented in Code~\ref{code:mps_qubit} initializes a 20-qubit circuit utilizing an MPS with a restricted bond dimension of 7.
The resulting quantum state is naturally returned in its MPS form---a collection of order-3 tensors, where each tensor corresponds to a physical site within the circuit.

\begin{lstlisting}[caption={Simulating a 20-mode photonic quantum circuit using MPS.}, label={code:mps_qumode}]
nmode = 20
cir = dq.QumodeCircuit(nmode, [1] * nmode, cutoff=4, backend='fock', basis=False, mps=True, chi=3)
for i in range(nmode - 1):
    cir.bs([i, i + 1])
mps_state = cir()
for state in mps_state:
    print(state.shape)
\end{lstlisting}

\noindent Output:
\begin{verbatim}
torch.Size([1, 4, 3])
torch.Size([3, 4, 3])
torch.Size([3, 4, 3])
......
torch.Size([3, 4, 1])
\end{verbatim}

Similarly, this tensor network backend is fully integrated into the \texttt{QumodeCircuit} class, as demonstrated in Code~\ref{code:mps_qumode}.

\subsection{Distributed Simulations}
To address the exponential memory requirements of simulating large-scale quantum systems, DeepQuantum realizes a distributed parallel computing architecture~\cite{jones2023distributed} that leverages PyTorch's native distributed communication package.
This powerful feature allows the quantum state vector to be partitioned and managed across multiple processes or GPUs, supporting both intra-node and inter-node parallelism, thereby enabling the simulation of quantum systems at a scale that exceeds the memory limits of individual devices.

The framework is designed to facilitate a seamless transition from a single-device simulation to a distributed one.
The primary change for the user is to replace the standard \texttt{QubitCircuit} class with its distributed counterpart, \texttt{DistributedQubitCircuit}.
The API for constructing and executing the quantum circuit remains syntactically identical.
This design philosophy allows developers to scale up their simulations with minimal code modification.

The example below illustrates the fundamental structure for a distributed simulation.
The script is launched via \texttt{torchrun}, which manages the instantiation of processes.
The \texttt{dq.setup\_distributed()} function handles the initialization of the communication backend (e.g., \verb|'nccl'| for CUDA GPU), while \texttt{dq.cleanup\_distributed()} ensures a clean shutdown.

\begin{lstlisting}[caption={A minimal example of a distributed quantum circuit simulation.
The script should be executed with the command \texttt{torchrun {-}{-}nproc\_per\_node=[NUM\_GPUS] script\_name.py}.}, label={code:dist}]
import deepquantum as dq
import torch

# Initialize the distributed environment
rank, world_size, local_rank = dq.setup_distributed(backend='nccl')
device = f'cuda:{local_rank}'
data = torch.arange(4, dtype=torch.float, device=device, requires_grad=True)
# Construct the circuit
cir = dq.DistributedQubitCircuit(4)
cir.rylayer(encode=True)
cir.cnot_ring()
cir.observable(0)
cir.observable(1, 'x')
cir.to(device)
# Execute the simulation
state = cir(data).amps
result = cir.measure(with_prob=True)
exp = cir.expectation().sum()
exp.backward()
# Print the results gathered at rank 0
if rank == 0:
    print(state)
    print(result)
    print(exp)
    print(data.grad)
# Clean up the distributed process group
dq.cleanup_distributed()
\end{lstlisting}

As shown in Code~\ref{code:dist}, the core logic of simulating the quantum circuit is completely unchanged.
\texttt{DistributedQubitCircuit} internally handles the complex task of partitioning the state vector and managing the inter-GPU communication required for gate operations that span across different state vector slices.
\texttt{measure} and \texttt{expectation} automatically aggregate results from all processes and make them available on the rank-0 process, abstracting away the communication overhead from the user.
The gradients of the expectation value with respect to differentiable inputs or trainable parameters are implemented through a custom \texttt{torch.autograd.Function} using the adjoint differentiation method.
This approach seamlessly integrates them into the whole computational graph.
Additionally, distributed simulations are also supported for the Fock backend based on state tensors via \texttt{DistributedQumodeCircuit}.
Our framework combines the power of large-scale distributed computing with the simplicity and ease of use characteristic of DeepQuantum.

\subsection{Applications}
\subsubsection{Simulation of Large Quantum Systems with MPS}
The transverse-field Ising model (TFIM) with nearest-neighbor interactions and periodic boundary conditions is governed by the Hamiltonian~\cite{Pfeuty1970}
\begin{equation}
    \hat{H}_{\mathrm{TFIM}} = \hat{H}_{ZZ} + \hat{H}_X = -J \sum_{\langle i,j \rangle} \hat{\sigma}_z^i \hat{\sigma}_z^j - h\sum_i \hat{\sigma}_x^i.
\end{equation}
Adopting natural units $\hbar=1$, the time evolution of this system can be represented by the unitary operator
\begin{equation}
    \ket{\psi(t)} = \hat{U}_{\mathrm{TFIM}}(t)\ket{\psi(0)} = \exp(-i\hat{H}_{\mathrm{TFIM}}t) \ket{\psi(0)}.
\end{equation}
To simulate this continuous-time evolution on a digital quantum processor, it is necessary to perform Trotterization:
\begin{equation}
    \begin{aligned}
        e^{-i\hat{H}_{\mathrm{TFIM}}t} &= \bqty{e^{-i\hat{H}_{\mathrm{TFIM}}\Delta t}}^N \\
        &= \bqty{e^{-i\hat{H}_X \frac{\Delta t}{2}} e^{-i\hat{H}_{ZZ} \Delta t} e^{-i\hat{H}_X \frac{\Delta t}{2}} + \mathcal{O}(\Delta t^3)}^N,
    \end{aligned}
\end{equation}
where $\Delta t=t/N$, with $N$ being the number of time steps.
The second equality represents the second-order Trotter-Suzuki decomposition~\cite{suzuki1992general, berry2007efficient}, which bounds the error per step to $\mathcal{O}(\Delta t^3)$~\cite{childs2021theory}.
In DeepQuantum, the evolutions under $\hat{H}_{ZZ}$ and $\hat{H}_{X}$ are straightforwardly implemented via the \texttt{QubitCircuit.rzz()} and \texttt{QubitCircuit.rx()} methods, respectively.

The evolution of the one-dimensional TFIM has been extensively studied using exact diagonalization.
The system can be mapped to free fermions via the Jordan-Wigner and Bogoliubov transformations~\cite{Pfeuty1970}, allowing for the analytical study of observable dynamics~\cite{barouch1970statistical1, barouch1971statistical2, barouch1971statistical3} under a quantum quench---a process where the initial state is not an eigenstate of the driving Hamiltonian.

\begin{lstlisting}[caption={Implementation of the TFIM quench simulation using MPS.}, label={code:mps_tfim}]
import deepquantum as dq

class TFIMSimulator:
    def __init__(self, nqubit, J, h):
        self.nqubit = nqubit
        self.J = J  # coupling strength
        self.h = h  # transverse field

    # Second-order Trotterization
    def trotter_step(self, circuit, dt):
        # X rotations (first half)
        for i in range(self.nqubit):
            circuit.rx(i, -self.h * dt)
        # Even ZZ interactions
        for i in range(0, self.nqubit - 1, 2):
            circuit.rzz([i, i + 1], -2 * self.J * dt)
        # Odd ZZ interactions
        for i in range(1, self.nqubit - 1, 2):
            circuit.rzz([i, i + 1], -2 * self.J * dt)
        # Periodic boundary condition
        circuit.rzz([self.nqubit - 1, 0], -2 * self.J * dt)
        # X rotations (second half)
        for i in range(self.nqubit):
            circuit.rx(i, -self.h * dt)
        return circuit

    # Evolve and measure TFIM
    def evolve_and_measure(self, max_steps, dt, chi):
        results = {}
        steps, magnetizations = [], []
        # Evolve step by step
        for step in range(max_steps + 1):
            cir = dq.QubitCircuit(self.nqubit, mps=True, chi=chi)
            for i in range(self.nqubit):
                cir.h(i)
            for _ in range(step):
                cir = self.trotter_step(cir, dt)
            for i in range(self.nqubit):
                cir.observable(i, basis='x')
            # Circuit execution
            cir()
            expectations = cir.expectation()
            avg_mag = sum(exp.real for exp in expectations) / self.nqubit
            steps.append(step)
            magnetizations.append(avg_mag)
        results = {
            'steps': steps,
            'magnetization': magnetizations,
            'times': [s * dt for s in steps]
        }
        return results
\end{lstlisting}

Here, we simulate the quantum quench of the TFIM using second-order Trotterization based on MPS.
The simulation employs $n=45$ qubits, a system size that far exceeds the limits of classical exact diagonalization.
The system is initialized in the product state $\ket{+}^{\otimes n}$.
The simulation parameters are set to $J=-1.0$, $h=-1.2$, $\Delta t=0.1$, and $N=50$.
We execute the simulation across a range of maximum bond dimensions $\chi \in \{8, 16, 32, 64, 128\}$ and track the time evolution of the average $X$-magnetization $\overline{\expval{\sigma_x}}$, defined as
\begin{equation}
    \overline{\expval{\sigma_x}} \equiv \frac{1}{n} \sum_{i=1}^n \expval{\hat{\sigma}_x^i},
\end{equation}
where $\hat{\sigma}_x^i$ is the Pauli-$X$ operator acting on the $i$-th qubit.
The core implementation is detailed in Code~\ref{code:mps_tfim}.

\begin{figure}[!htbp]
    \centering
    \includegraphics[width=\linewidth]{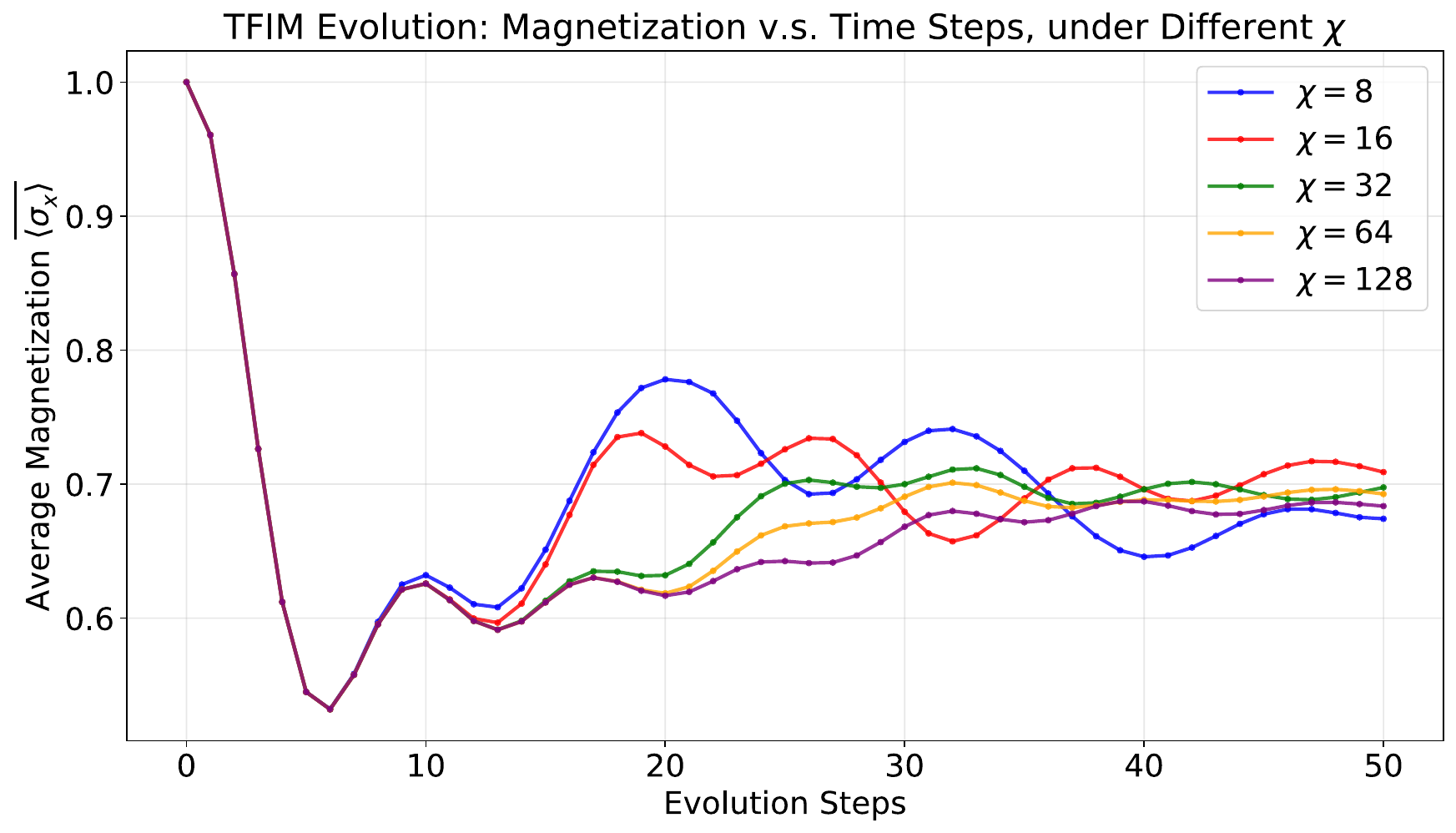}
    \caption{MPS Simulation of the TFIM quench dynamics for a 45-qubit system.
    The time evolution of the average $X$-magnetization is computed under different maximum bond dimensions $\chi \in \{8, 16, 32, 64, 128\}$.
    The deviation of low-$\chi$ curves at later stages illustrates the accumulation of truncation errors.}
    \label{fig:mps_tfim}
\end{figure}

The numerical results are shown in Fig.~\ref{fig:mps_tfim}.
Following the quench, the trajectory calculated with the largest bond dimension ($\chi=128$) reveals that the observable $\overline{\expval{\sigma_x}}$ relaxes over time, with its expectation value progressively approaching a stationary limit.
This phenomenon arises from the dephasing between post-quench eigenstates, causing their characteristic oscillations to destructively interfere and cancel out in the long-time limit~\cite{barouch1971statistical2, barouch1971statistical3}.
Furthermore, the comparison across different $\chi$ values highlights the necessity of sufficient tensor expressivity: simulations with lower bond dimensions (e.g., $\chi=8, 16$) deviate significantly from the large-$\chi$ reference trajectory after 15 steps due to truncation errors, whereas $\chi=128$ effectively captures the long-term stabilization.

\subsubsection{Distributed Quantum Fourier Transform}
To further demonstrate the large-scale simulation capabilities of our framework, we implement a distributed quantum Fourier transform (QFT), a critical subroutine in numerous quantum algorithms including Shor's algorithm.
Because the QFT circuit contains a dense pattern of long-range controlled-phase operations across many qubits, it serves as a rigorous benchmark for evaluating the computational throughput and memory management of a quantum simulator.

\begin{lstlisting}[caption={Object-oriented implementation of a distributed QFT circuit.}, label={code:dist_qft}]
import deepquantum as dq
import torch

class DistQFT(dq.DistributedQubitCircuit):
    def __init__(self, nqubit, reverse=False):
        super().__init__(nqubit=nqubit)
        self.wires = list(range(nqubit))
        self.minmax = [0, nqubit - 1]
        for i in self.wires:
            self.qft_block(i)
        if not reverse:
            for i in range(len(self.wires) // 2):
                self.swap([self.wires[i], self.wires[-1 - i]])

    def qft_block(self, n):
        self.h(n)
        k = 2
        for i in range(n, self.minmax[1]):
            self.cp(i + 1, n, torch.pi / 2 ** (k - 1))
            k += 1
\end{lstlisting}

As shown in Code~\ref{code:dist_qft}, the core logic is encapsulated within the \texttt{qft\_block} method, which iteratively applies a Hadamard gate followed by a cascading series of controlled-phase rotations.
This circuit structure can be expressed concisely using DeepQuantum's intuitive API.

We benchmarked the performance of the distributed QFT simulation using a cluster of 8 NVIDIA A100 GPUs, scaling the system size up to 33 qubits.
We recorded the execution time for a single forward pass of the complete circuit, followed by a state measurement operation.
To ensure statistical consistency, the results for each configuration represent the average over 10 independent runs, excluding the initial warm-up pass.

\begin{figure}[!htbp]
    \centering
    \includegraphics[width=\linewidth]{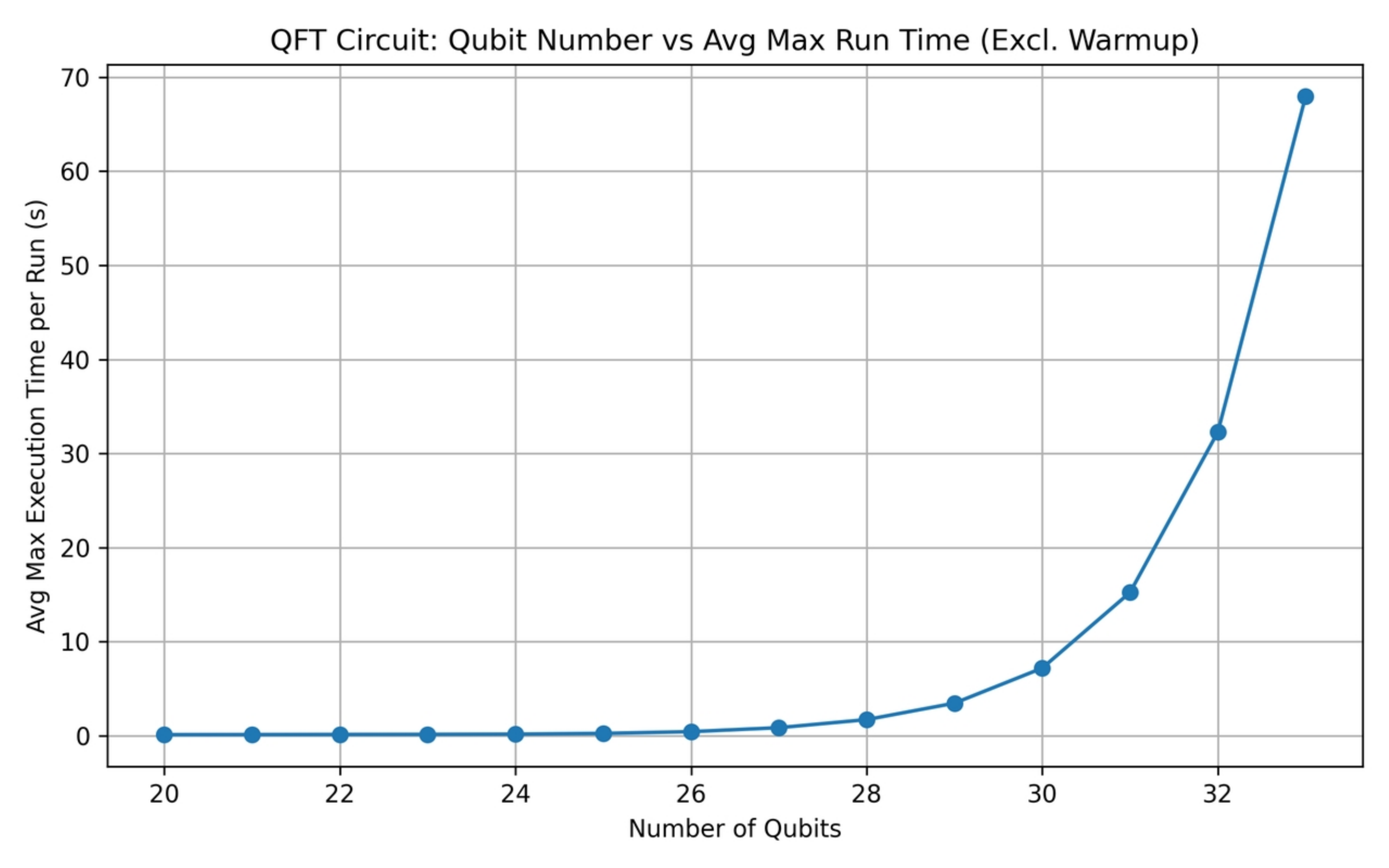}
    \caption{Average execution time for the distributed QFT simulation on 8 NVIDIA A100 GPUs as a function of the number of qubits.
    The exponential scaling of simulation time directly reflects the exponential growth of the Hilbert space inherent in full state-vector simulations.}
    \label{fig:qft_performance}
\end{figure}

The benchmark results, summarized in Fig.~\ref{fig:qft_performance}, demonstrate DeepQuantum's capacity to perform large distributed state-vector simulations within a highly competitive timeframe.
Notably, a 30-qubit QFT simulation completes in approximately 7.2 seconds, while the 33-qubit simulation is accomplished in just over one minute.

\begin{figure}[!htbp]
    \centering
    \includegraphics[width=\linewidth]{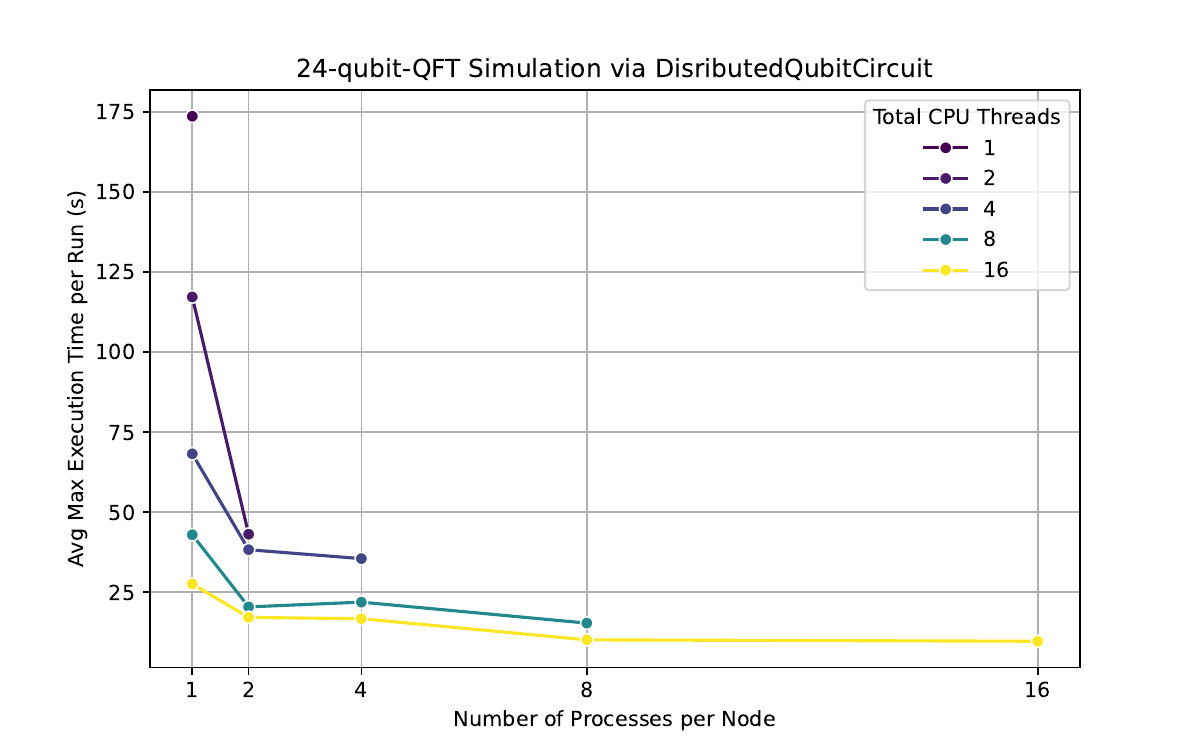}
    \caption{Average execution time of a 24-qubit QFT simulation in a multi-core CPU environment, evaluated under various configurations of PyTorch distributed processes and OpenMP threads.}
    \label{fig:qft_parallel}
\end{figure}

Furthermore, the distributed computational workload can be significantly accelerated on CPUs through a hybrid parallelization strategy.
Specifically, PyTorch's distributed framework manages multi-process execution, while OpenMP is leveraged for intra-process parallelization of compute-intensive operations, thereby efficiently utilizing all available CPU cores.
To verify this, we simulated a 24-qubit QFT in a multi-core CPU environment.
The profiling results in Fig.~\ref{fig:qft_parallel} indicate that selecting an optimal configuration of distributed processes and threading hyperparameters yields a multi-fold acceleration compared to conventional serial execution.

Ultimately, these performance metrics underscore the efficacy of DeepQuantum's distributed architecture in tackling computationally demanding, large-scale quantum simulations that lie beyond the reach of conventional single-device methods.

\section{Conclusion}
\label{sec:conclusion}
In this work, we have introduced DeepQuantum, a comprehensive software platform tailored for quantum machine learning and photonic quantum computing.
By building directly upon PyTorch, DeepQuantum enables seamless integration of quantum components as native elements within dynamic computational graphs, allowing researchers to fully utilize PyTorch's automatic differentiation, optimization tools, and rich machine learning ecosystem with minimal overhead.
Uniquely, the platform provides a unified programming interface that integrates three fundamental paradigms of quantum computation: the conventional qubit-based model, the photonic quantum models, and the MBQC model.
This integrated approach, supported by scalable tensor network simulations and distributed architecture, empowers users to explore a much broader landscape of hybrid algorithms and accelerate the discovery of AI-driven solutions for practical applications.
Through various benchmarks and use cases, we have demonstrated DeepQuantum's capability in variational algorithm implementation, photonic circuit simulation, and large-scale quantum computations, establishing it as a versatile and practical tool for advancing quantum information processing.

Looking forward, we plan to expand DeepQuantum's library with more advanced photonic components, enhanced noise modeling capabilities, and improved hardware integration interfaces.
Further optimization of computational performance and extended support for hybrid quantum-classical workflows are also key priorities.
These developments will broaden the platform's applicability, serving as a robust foundation for both academic research and industrial innovation in the evolving domains of AI-assisted QC and quantum-enhanced AI.

\section*{Acknowledgments}
This research is supported by the National Key R\&D Program of China (Grants No. 2024YFB4504005, No. 2024YFA1409300); National Natural Science Foundation of China (NSFC) (Grants No. 62235012, No. 12304342, No. 12574549, No. 12574542, No. 125B1033); Innovation Program for Quantum Science and Technology (Grants No. 2021ZD0301500, and No. 2021ZD0300700); Science and Technology Commission of Shanghai Municipality (STCSM) (Grants No. 2019SHZDZX01, No. 24ZR1438700, No. 24ZR1430700 and No. 24LZ1401500); Startup Fund for Young Faculty at SJTU (SFYF at SJTU) (Grants No. 24X010502876 and No. 24X010500170); Frontier Technologies R\&D Program of Jiangsu (Grant No. SBF20250000094); Zhiyuan Future Scholar Program (Grant No. ZIRC2024-06).
X.-M.J. acknowledges additional support from a Shanghai talent program and support from Zhiyuan Innovative Research Center of Shanghai Jiao Tong University.
H. T. acknowledges additional support from Yangyang Development Fund.

\bibliographystyle{naturemag}
\bibliography{ref}


\end{document}